\documentclass[11pt]{article}
\pdfoutput=1
\usepackage{amsmath,amssymb,amsthm,amsxtra,overpic,bbm,bm,epsfig}
\usepackage{color,subfigure}
\usepackage{soul}
\usepackage[dvipsnames]{xcolor}

\def\thefootnote{\fnsymbol{footnote}}
\allowdisplaybreaks[4]

\newcommand{\bq}{\begin{eqnarray}}
\newcommand{\nq}{\end{eqnarray}}
\newcommand{\tildex}{\overset{ \text {\tiny $(\sim)$} }}

\begin{document}

\begin{flushright} 
IPPP/17/86  \\
\end{flushright} 

\begin{center}
{\Large\bf Effective alignments as building blocks of flavour models} \\

\end{center} 
\vspace{0.2cm}

\begin{center}
{\bf Ivo de Medeiros Varzielas$^1$}\footnote{Email: \tt ivo.de@udo.edu}
{\bf Thomas Neder$^2$}\footnote{Email: \tt neder@ific.uv.es}
and 
{\bf Ye-Ling Zhou$^3$}\footnote{Email: \tt ye-ling.zhou@durham.ac.uk}
\\\vspace{5mm}
{$^1$CFTP, Departamento de F\'{\i}sica, Instituto Superior T\'{e}cnico,}\\
Universidade de Lisboa,
Avenida Rovisco Pais 1, 1049 Lisboa, Portugal \\
{$^2$ AHEP  Group,  Instituto  de  F\'{\i}sica  Corpuscular (IFIC) --- C.S.I.C./Universitat  de  Val\`{e}ncia} 
\\
{$^3$ Institute for Particle Physics Phenomenology, Department of Physics, \\ Durham University, Durham DH1 3LE, United Kingdom} 
\\
\end{center}

\vspace{1.5cm} 

\begin{abstract} 

Flavour models typically rely on flavons - scalars that break the family symmetry by acquiring vacuum expectation values in specific directions. We develop the idea of effective alignments, i.e.\ cases where the contractions of multiple flavons give rise to directions that are hard or impossible to obtain directly by breaking the family symmetry.
Focusing on the example where the symmetry is $S_4$, we list the effective alignments that can be obtained from flavons vacuum expectation values that arise naturally from $S_4$. Using those effective alignments as building blocks, it is possible to construct flavour models, for example by using the effective alignments in constrained sequential dominance models. We illustrate how to obtain several of the mixing schemes in the literature, and explicitly construct renormalizable models for three viable cases, two of which lead to trimaximal mixing scenarios.

\end{abstract}
\begin{flushleft}
\hspace{0.8cm} PACS number(s): 14.60.Pq, 11.30.Hv, 12.60.Fr \\
\hspace{0.8cm} Keywords: Lepton flavour mixing, cross couplings, flavour symmetry
\end{flushleft}

\def\thefootnote{\arabic{footnote}}
\setcounter{footnote}{0}

\newpage

%\tableofcontents

\section{Introduction}

For decades, non-Abelian discrete flavour symmetries have been widely used in models of lepton flavour mixing (see some reviews, e.g., \cite{Altarelli:2010gt, King:2013eh, King:2014nza}). The basic idea is to assume a non-Abelian discrete symmetry in flavour space and introduce scalars, called flavons, which obtain non-trivial VEVs that break this flavour symmetry such that specific flavour structures in the lepton mass matrices, and thus also in the mixing matrix, are realised.

One great achievement of flavour symmetries is the prediction of mixing patterns where the mixing matrix is completely fixed by the symmetry (up to permutations of rows and columns), e.g., predicting tri-bimaximal mixing \cite{Harrison:2002er, Xing:2002sw, Harrison:2002kp, He:2003rm} in $A_4$ \cite{Altarelli:2005yp, Altarelli:2005yx} and $S_4$ \cite{Lam:2008sh} and democratic mixing \cite{Fritzsch:1995dj, Fritzsch:1998xs} in $S_4$ \cite{Ding:2013eca} and bimaximal mixing \cite{Vissani:1997pa, Barger:1998ta} in $S_4$ \cite{Feruglio:2012cw}, or the mixing in \cite{Toorop:2011jn}. Usually, these patterns are realised in the so-called direct or semi-direct approaches \cite{King:2009ap}, where in semi-direct approaches the mixing matrix is determined up to a rotation left unfixed by the symmetry. In these approaches, some residual symmetries, e.g., $Z_n$ in the charged lepton sector, $Z_2$ or $Z_2 \times Z_2'$ in the neutrino sector (if neutrinos are Majorana particles), are preserved after the breaking of the full symmetry, and the mixing mainly results from the misalignment of the different residual symmetries in charged lepton and neutrino sectors. One advantage of these approaches is a connection between the flavour mixing structure and group structure: one can in principle predict the mixing matrix based on the assumed flavour symmetry without going into the  details of flavour models. The semi-direct approach has also been combined with general CP symmetries, resulting in the prediction of all mixing parameters (either constant or dependent upon one single free parameter) \cite{Feruglio:2012cw, Holthausen:2012dk}. However, constructing flavour models with respect to residual symmetries is not straightforward. In order to preserve a residual symmetry, one has to carefully construct the flavon potential (or superpotential in the framework of supersymmetry) to include the desired couplings and avoid unnecessary couplings in the Lagrangian. In order to explain the observed large reactor angle and the hinted-at maximal Dirac-type CP violation, corrections with special directions or non-trivial phases are required in the flavon VEVs. These requirements increase the complexity of models.

The so-called indirect approach provides an alternative way of attempting to understand flavour mixing. In this approach, the underlying flavour symmetry is completely broken by flavon VEVs and the flavour mixing is directly generated by flavon VEVs, in a way which will be explained in the following. A typical example is constrained sequential dominance (CSD), which provides a framework which allows the mixing angles and phases to be accurately predicted in terms of relatively few input parameters (see \cite{King:2005bj} and e.g.\ \cite{Antusch:2011ic, Bjorkeroth:2014vha, Bjorkeroth:2015ora, Bjorkeroth:2015uou, Bjorkeroth:2017ybg}). One important but quite generic assumption of CSD is that right-handed (RH) neutrinos $N_i$ (for $i=1,2,3$) are assumed to be flavour singlets with masses $M_{N_1}, M_{N_2}, M_{N_3}$ \footnote{$M_{N_1}$, $M_{N_2}$, and $M_{N_3}$ are denoted as $M_\text{atm}$, $M_\text{sol}$ and $M_\text{dec}$, respectively \cite{King:2013xba}, although we don't follow this notation here as we leave the ordering of the masses general.}, and that through flavon VEVs the Yukawa couplings lead to the Dirac neutrino mass term $M_D \approx \{Y_1, Y_2, Y_3\}$, and the active neutrino mass matrix is given by
\bq
M_\nu \approx \sum_i \frac{v^2}{M_{N_i}} Y_i Y_i^T \,.
\nq
Here, each $Y_i$ is a $3\times 1$ (column) vector and $v$ is the SM-Higgs VEV. In CSD$n$ models,  \cite{Bjorkeroth:2014vha}, $Y_1 \propto (0,1,-1)^T$ and $Y_2 \propto (1,n,2-n)^T$ are respectively responsible for atmospheric angle $\theta_{23}$ and solar angle $\theta_{12}$ (note that the placement of the signs in these directions is convention dependent). The littlest seesaw model, \cite{Ballett:2016yod}, i.e., CSD3 with $N_3$ decoupled, introduces only three real parameters, but can fit three mixing angles and two mass-square differences. Although it is predictive, the theoretical justification of $Y_2 \propto (1,3,-1)^T$ from symmetry has only been obtained through relatively complicated alignment mechanisms relying on superpotential terms \cite{Bjorkeroth:2015ora, Bjorkeroth:2015uou}. 
For the CSD2 model, the directions (1,0,2) or (1,2,0) have been realised through slightly simpler alignments \cite{Antusch:2011ic}.

The main aim of this paper is to explore models of neutrino masses in which in higher-order operators flavon VEVs combine to new, effective alignments (EAs), the latter giving rise to fermion mass terms.
For this we consider how EAs arise and how they can be used in models.
By EAs we refer to directions that are obtained from starting with the simple flavon vacuum expectation value (VEV) directions that are obtained from the spontaneous breaking of the symmetry and then contracting multiples of the simple directions according to the rules of the symmetry group.
We use the group $S_4$ as an example, although the strategies and conclusions generalise for other groups.

We start by listing the possible VEV alignments that are allowed by the potential of one scalar triplet of the flavour group (which coincide with the results of \cite{Ivanov:2014doa, deMedeirosVarzielas:2017glw, deMedeirosVarzielas:2017ote}) and which serve as the building blocks of EAs.

We then consider multiple flavons, and discuss how in higher-order operators, two or more flavons can be combined to form EAs and how these can be implemented to account for leptonic mixing, particularly using sequential dominance \cite{King:1998jw, King:1999cm, King:1999mb, King:2002nf} in Constrained Sequential Dominance (CSD) models (see e.g.\ \cite{Antusch:2011ic, Bjorkeroth:2014vha, Bjorkeroth:2015ora, Bjorkeroth:2015uou, Bjorkeroth:2017ybg} for recent examples, and references therein). 
In addition, we extend the discussion from previous CSD models by not specifying the hierarchy between $M_1$, $M_2$ and $M_3$.
Although the idea of EAs is not new \cite{Varzielas:2015aua}, it remains unexplored as implementation at the non-renormalizable level leads to difficulties where typically a desired EA cannot be separated from other undesired EAs which at best reduce the predictivity and at worst spoil the fermion masses and mixing. Here we demonstrate specific UV completions that forbid most contractions of the VEVs that are in principle allowed at the effective level (and which would lead to undesirable contributions). The idea of using fermion messengers was originally used in the Froggatt-Nielsen mechanism \cite{Froggatt:1978nt}. It has been widely used in non-Abelian discrete flavour models, many of which implement specific UV completions to make the models more predictive (see e.g. \cite{Varzielas:2010mp, Varzielas:2012ai, Ding:2013hpa, Ding:2014hva}). In our work this is also the case, as by assuming right-handed neutrinos to be singlets in the flavour space, we obtain a simple one-to-one correspondence between the UV completion and a single flavon contraction.

Assuming the single triplet alignments of $S_4$ for multiple flavons, we present a list of possible EAs for $S_4$ obtained by combining up to three flavon directions.
We then demonstrate the EA strategy by using some of the EAs of $S_4$ to build three phenomenologically viable UV complete models, which include one EA that arises from the combination of 3 distinct VEVs in a specific ordering.
These models lead to specific neutrino mass matrix structures that had not been studied so far, which we briefly discuss their phenomenology with respect to the leptonic mass and mixing parameters (including their respective prediction for the Dirac CP phase, and for neutrinoless double beta decay).

The outline of the paper is as follows. In Section \ref{sec:VEV} we consider the $S_4$ symmetry and list the directions of the VEVs in two bases of interest. In Section \ref{sec:Blocks} we establish the building block for constructing models: the effective couplings arising from UV completions are described in a general, model-independent way in Section \ref{sec:UV_eff}, the EAs obtained from the VEVs (illustrated with the $S_4$ VEVs) in Section \ref{sec:eff_dir}, and the lepton mixing in CSD models in Section \ref{sec:lep_mix}. In Section \ref{sec:CSD2 models} we combine these aspects together and build 3 example models with the EAs of $S_4$ in the CSD framework, and present their phenomenology in terms of the respective masses and mixing parameters.
In Section \ref{sec:cross_coupling} we discuss cross-couplings between flavons and how they may affect the previous results.

\section{Flavon vacuum alignments \label{sec:VEV}}

Consider the potential of a complex triplet flavon $\varphi=(\varphi_1, \varphi_2, \varphi_3)^T$ transforming as a $\mathbf{3}$ or $\mathbf{3^\prime}$ of $S_4$. To simplify the calculation, we work in the first basis listed in Appendix \ref{app:S4}. In this basis, $\varphi^\dagger=\tilde{\varphi}=(\varphi_1^*, \varphi_2^*, \varphi_3^*)^T$ is also a triplet $\mathbf{3}$ ($\mathbf{3^\prime}$) of $S_4$ \footnote{In our notation, $\tilde{\varphi}$ means that $\tilde{\varphi}$ is also a triplet of $S_4$. $\tilde{H}_\alpha$ (which appears below) is also an $SU(2)$ doublet, whereas $H^{\dagger \alpha}$ doesn't transform exactly as a doublet, so one uses $\epsilon_{\alpha \beta}$.}. To forbid unnecessary couplings such as $\varphi^3$ or $\varphi^4$, we introduce a global $U(1)_F$ symmetry, which makes the potential analogous to the potential of a $S_4$ triplet which is also an $SU(2)_L$ doublet \cite{Ivanov:2014doa, deMedeirosVarzielas:2017glw}. Also due to the additional $U(1)_F$ symmetry, we have the same expression for the potential for both $\varphi$ being a {\bf 3} or a {\bf 3'} of $S_4$. In fact, as this $U(1)_F$ is broken by the flavon VEVs, the models would have a massless Goldstone boson, in conflict with phenomenology. To avoid this, one can either gauge the continuous symmetry and have it broken at sufficiently high scale to avoid experimental bounds, or consider a discrete subgroup $Z_N^F$ with sufficiently high $N$ to avoid accidental terms up to a certain order, but in that case $N$ should also not be too large to prevent an accidental global $U(1)$ symmetry (in which case the Goldstone boson again arises).
There are five Kronecker products involving one triplet under $S_4$ with an additional $U(1)_F$ charge that produce singlets, and adding those up produces the potential
\bq 
\hspace{-5mm}V(\varphi)&=&\mu^2_\varphi (\tilde{\varphi} \varphi)_\mathbf{1} + f_1 \big( (\tilde{\varphi} \varphi)_\mathbf{1}\big)^2 + f_2 \big( (\tilde{\varphi} \varphi)_{\mathbf{2}} (\tilde{\varphi} \varphi)_{\mathbf{2}} \big)_\mathbf{1} + f_3 \big( (\tilde{\varphi} \varphi)_{\mathbf{3}} (\tilde{\varphi} \varphi)_{\mathbf{3}} \big)_\mathbf{1} + f_4 \big( (\tilde{\varphi} \varphi)_{\mathbf{3'}} (\tilde{\varphi} \varphi)_{\mathbf{3'}} \big)_\mathbf{1}  \,,
\label{eq:potential}
\nq
where all the coefficients $\mu_\varphi$ and $f_i$ are real. However, these invariants are not independent\footnote{The potentials has an additional symmetry under exchange of the positions of $\varphi$  with itself and similarly under $\tilde{\varphi}$, so-called plethysms \cite{Fonseca:2017lem}.}, and the potential can be simplified to the following form
\bq
V(\varphi)&=& \mu^2_\varphi I_1 + g_1 I_1^2 + g_2 I_2 + \frac{1}{2}g_3( I_3 + \text{h.c.})
\label{eq:Vphi}
\nq
with
\bq
I_1 &=& |\varphi_1|^2 + |\varphi_2|^2 + |\varphi_3|^2\,,\nonumber\\
I_2 &=& |\varphi_1\varphi_2|^2 + |\varphi_2\varphi_3|^2 + |\varphi_3\varphi_1|^2\,,\nonumber\\
I_3 &=& (\varphi_1^*\varphi_2)^2 + (\varphi_2^*\varphi_3)^2 + (\varphi_3^*\varphi_1)^2\,,
\nq
and
\bq
g_1 &=& f_1+f_2\,, \nonumber\\
g_2 &=& f_3+f_4-3f_2\,, \nonumber\\
g_3 &=& f_3+f_4\,.
\nq
In order to achieve a non-trivial and stable vacuum, we require the quadratic term to be negative-definite and the quartic terms to be positive-definite, which lead to $\mu_\varphi^2<0$, and $g_1>0$, $3g_1+g_2>|g_3|$, respectively.

The flavour symmetry is spontaneously broken after the flavon $\varphi$ gets a VEV $\langle \varphi_i \rangle=v_i e^{i\alpha_i}/\sqrt{2}$, where $v$ and $\alpha$ are real, $v>0$, and $0\leqslant\alpha< 360^\circ$. The VEVs are well-known in the literature and have been presented in e.g.\ \cite{Ivanov:2014doa, deMedeirosVarzielas:2017glw}, with their CP properties analysed in \cite{deMedeirosVarzielas:2017ote}. The classical minimisation is given for completeness in Appendix \ref{app:min}.

In total, there are four classes of candidates of the $\varphi$ VEV, they are given by
\bq
\langle \varphi \rangle_\text{I} &=& \left\{
\begin{pmatrix}1 \\ 0 \\ 0 \end{pmatrix} \,,\quad
\begin{pmatrix}0 \\ 1 \\ 0 \end{pmatrix} \,,\quad
\begin{pmatrix}0 \\ 0 \\ 1 \end{pmatrix} 
\right\} v_\text{I}\,;\nonumber\\
\langle \varphi \rangle_\text{II} &=& \left\{ 
\begin{pmatrix}0 \\ e^{+ i \pi/4} \\  e^{- i \pi/4} \end{pmatrix}, 
\begin{pmatrix}0 \\ e^{- i \pi/4} \\  e^{+ i \pi/4} \end{pmatrix},
\begin{pmatrix} e^{- i \pi/4} \\ 0 \\ e^{+ i \pi/4} \end{pmatrix},
\begin{pmatrix} e^{+ i \pi/4} \\ 0 \\ e^{- i \pi/4} \end{pmatrix},
\begin{pmatrix} e^{+ i \pi/4} \\ e^{- i \pi/4} \\ 0 \end{pmatrix},
\begin{pmatrix} e^{- i \pi/4} \\ e^{+ i \pi/4} \\ 0 \end{pmatrix}
\right\} \frac{v_\text{II}}{\sqrt{2}}\,;\nonumber\\
\langle \varphi \rangle_\text{III} &=& \left\{ 
\begin{pmatrix}1 \\ 1 \\ 1 \end{pmatrix},
\begin{pmatrix}-1 \\ 1 \\ 1 \end{pmatrix},
\begin{pmatrix}1 \\ -1 \\ 1 \end{pmatrix},
\begin{pmatrix}1 \\ 1 \\ -1 \end{pmatrix} 
\right\} \frac{v_\text{III}}{\sqrt{3}}\,;\nonumber\\
\langle \varphi \rangle_{\text{III}'} &=& \left\{ 
\begin{pmatrix}1 \\ \omega \\ \omega^2 \end{pmatrix},
\begin{pmatrix}1 \\ \omega^2 \\ \omega \end{pmatrix}, 
\begin{pmatrix}1 \\ -\omega \\ -\omega^2 \end{pmatrix},
\begin{pmatrix}1 \\ -\omega^2 \\ -\omega \end{pmatrix}, 
\begin{pmatrix}1 \\ -\omega \\ \omega^2 \end{pmatrix},
\begin{pmatrix}1 \\ -\omega^2 \\ \omega \end{pmatrix},
\begin{pmatrix}1 \\ \omega \\ -\omega^2 \end{pmatrix},
\begin{pmatrix}1 \\ \omega^2 \\ -\omega \end{pmatrix} 
 \right\} \frac{v_{\text{III}'}}{\sqrt{3}}\,.
\nq
where $v_{\text{I},\text{II},\text{III},\text{III}'}$ can always be chosen to be real and positive parameters through a phase redefinition of the flavon $\varphi$. Which VEV $\varphi$ achieves depends on the relations of coefficients $g_i$, which are also summarised in Appendix B.

The basis we used so far is particularly useful to study the potential and VEVs, and was the basis used throughout in \cite{Ivanov:2014doa, deMedeirosVarzielas:2017glw}.
Now we re-express these VEVs in a second basis, which is particularly convenient for building lepton flavour models. For the models presented in Section \ref{sec:lep_mix}, this second basis corresponds to the basis where the charged lepton mass matrix is diagonal. The triplet representation in the second basis is obtained through the transformation $\varphi \to U_\omega \varphi$, where $\omega=e^{i2\pi/3}$ and 
\bq
U_\omega = \frac{1}{\sqrt{3}}\left(
\begin{array}{ccc}
 1 & 1 & 1 \\
 1 & \omega ^2 & \omega  \\
 1 & \omega  & \omega ^2 \\
\end{array}
\right)\,.
\nq
In this basis, if we have a triplet, we can build another triplet:
\bq
\varphi=(\varphi_1, \varphi_2, \varphi_3)^T \sim \mathbf{3} \to  
\tilde{\varphi}=(\varphi_1^*, \varphi_3^*, \varphi_2^*) \sim \mathbf{3}
\label{eq:triplet_swap}
\nq
(note the swap in the component position) and conversely, if $\varphi$ is a $\mathbf{3'}$ then the corresponding $\tilde{\varphi}$ is also a $\mathbf{3'}$. The classes of flavon VEV candidates are given in the second basis by
\bq
\langle \varphi \rangle_\text{I} &=& \left\{
\begin{pmatrix}1 \\ 1 \\ 1 \end{pmatrix}\,,~
\begin{pmatrix}1 \\ \omega^2 \\ \omega \end{pmatrix}\,,~
\begin{pmatrix}1 \\ \omega \\ \omega^2 \end{pmatrix}
\right\} \frac{v_\text{I}}{\sqrt{3}} \,, \nonumber\\
\langle \varphi \rangle_\text{II} &=& \left\{ 
\begin{pmatrix} \frac{1}{\sqrt{3}} \\ \frac{1}{3+\sqrt{3}} \\ \frac{-1}{3-\sqrt{3}} \end{pmatrix}, 
\begin{pmatrix} \frac{1}{\sqrt{3}} \\ \frac{-1}{3-\sqrt{3}} \\ \frac{1}{3+\sqrt{3}} \end{pmatrix}, 
\begin{pmatrix} \frac{1}{\sqrt{3}} \\ \frac{\omega^2}{3+\sqrt{3}} \\ \frac{-\omega }{3-\sqrt{3}} \end{pmatrix}, 
\begin{pmatrix} \frac{1}{\sqrt{3}} \\ \frac{-\omega^2}{3-\sqrt{3}} \\ \frac{\omega }{3+\sqrt{3}} \end{pmatrix}, 
\begin{pmatrix} \frac{1}{\sqrt{3}} \\ \frac{\omega }{3+\sqrt{3}} \\ \frac{-\omega^2}{3-\sqrt{3}} \end{pmatrix}, 
\begin{pmatrix} \frac{1}{\sqrt{3}} \\ \frac{-\omega }{3-\sqrt{3}} \\ \frac{\omega^2}{3+\sqrt{3}} \end{pmatrix} 
\right\} v_\text{II}\,, \nonumber\\
\langle \varphi \rangle_\text{III} &=& \left\{ \begin{pmatrix} 1 \\ 0 \\0 \end{pmatrix}\,, ~
\begin{pmatrix} \frac{1}{3} \\ -\frac{2}{3} \\ -\frac{2}{3} \end{pmatrix}, ~
\begin{pmatrix} \frac{1}{3} \\ -\frac{2}{3}\omega^2 \\ -\frac{2}{3}\omega \end{pmatrix}, ~
\begin{pmatrix} \frac{1}{3} \\ -\frac{2}{3}\omega \\ -\frac{2}{3}\omega^2 \end{pmatrix}
\right\} v_\text{III}\,, \nonumber\\
\langle \varphi \rangle_{\text{III}'} &=& \left\{ 
\begin{pmatrix} 0 \\ 0 \\ 1 \end{pmatrix},
\begin{pmatrix} 0 \\ 1 \\ 0 \end{pmatrix},
\begin{pmatrix} \frac{2}{3} \\ \frac{2}{3} \\ -\frac{1}{3} \end{pmatrix},
\begin{pmatrix} \frac{2}{3} \\ -\frac{1}{3} \\ \frac{2}{3} \end{pmatrix},
\begin{pmatrix} -\frac{2}{3}\omega \\ -\frac{2}{3}\omega^2 \\ \frac{1}{3} \end{pmatrix},
\begin{pmatrix} -\frac{2}{3}\omega^2 \\ \frac{1}{3} \\ -\frac{2}{3}\omega \end{pmatrix},
\begin{pmatrix} -\frac{2}{3}\omega^2 \\ -\frac{2}{3}\omega \\ \frac{1}{3} \end{pmatrix},
\begin{pmatrix} -\frac{2}{3}\omega \\ \frac{1}{3} \\ -\frac{2}{3}\omega^2 \end{pmatrix}
\right\} v_{\text{III}'}\,. 
\label{eq:flavon_vevs}
\nq

\section{Building blocks for constructing models \label{sec:Blocks}}

\subsection{From UV completion to effective couplings \label{sec:UV_eff}}

There are four classes of VEVs for the flavon triplet after $S_4$ is spontaneously broken. We will now explain how to use these VEVs as building blocks for flavour model building in a general way. In order to connect these VEVs with flavour mixing, consider first a UV-complete theory\footnote{Specific examples are presented in Section \ref{sec:lep_mix}.} with $M$ copies of flavons $\Phi_1, \Phi_2,\cdots, \Phi_M$ and vector-like Dirac fermion mediators with heavy masses, $F_1, F_2,\cdots, F_M$ with their respective $\overline{F_I}$ in the usual sense, $\overline{F}_I \equiv F_I^\dagger \gamma_0$ (as for the other fermions). Here $\Phi_I\sim \mathbf{r}_{\Phi_I}$ and $F_I\sim \mathbf{r}_{F_I}\equiv\mathbf{r}_I$. 
Assume the following renormalizable Lagrangian terms,
\begin{eqnarray}
-\mathcal{L}_\text{UV} = y \tilde{H} \overline{\ell_L}F_M +  \sum_{I=2}^{M} y_I \Phi_I \overline{F_{I}} F_{I-1} + y_1 \Phi_1 \overline{F_1} N + \sum_{I=1}^{M}M_I (\overline{F_I} F_I)_\mathbf{1} + \text{h.c.}\,.
\label{eq:UV}
\end{eqnarray}
Here, the notation is such that particle identities are not specified: $F_I$ could be identical with $F_J$ or $\Phi_{I'}$ identical with $\Phi_{J'}$ for $I \neq J$ and $I' \neq J'$, respectively. The important aspect is that couplings exist which allow the procedure outlined in the remainder of this subsection.
We assume the $S_4$ representations are for the Higgs $H\sim \mathbf{1}$, for the lepton doublet $\ell_L\sim \mathbf{3}$ and for the right-handed neutrino $N\sim \mathbf{1}$ or $\mathbf{1'}$. To form an invariant  term $\tilde{H} (\overline{\ell_L}F_M)_\mathbf{1}$ of $S_4$, the representation of $F_M$ must be the same as that of $\ell_L$, i.e., $\mathbf{r}_M=\mathbf{3}$. 
Due to the property in Eqs.~\eqref{eq:invariance} and \eqref{eq:invariance_p} (in Appendix \ref{app:S4}), $\Phi_I \overline{F_I} F_{I+1} $ is identical with $\Big((\Phi_I \overline{F_{I}})_{\mathbf{r}_{I-1}} F_{I-1} \Big)_{\mathbf{1}}$, $\Big(\Phi_I (\overline{F_{I}} F_{I-1})_{\mathbf{r}_{\Phi_I}} \Big)_{\mathbf{1}}$ and $\Big(\overline{F_{I}} (F_{I-1} \Phi_I)_{\mathbf{r}_{I}} \Big)_{\mathbf{1}}$. To form a trivial singlet $(\Phi_1 \overline{F_1})_\mathbf{1} N$ (if $N \sim \mathbf{1}$) or $(\Phi_1 \overline{F_1})_\mathbf{1'} N$ (if $N \sim \mathbf{1'}$), the representations of $F_1$ and $\Phi_1$ are also related with each other. 

\begin{figure}
\begin{center}
\includegraphics[scale=0.7]{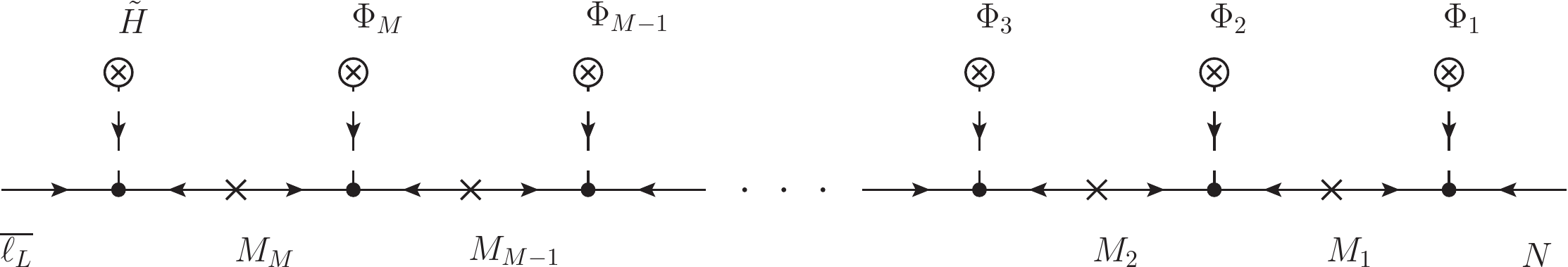}
\caption{General diagram of UV completion of the flavon contraction by introducing vector-like fermion $F_I$ (for $I=1,2,...,M$) with mass $M_I$. The relative renormalizable Lagrangian is given in Eq.~\eqref{eq:UV}.}
\label{fig:general}
\end{center}
\end{figure}

\begin{figure}
\begin{center}
\includegraphics[scale=1.0]{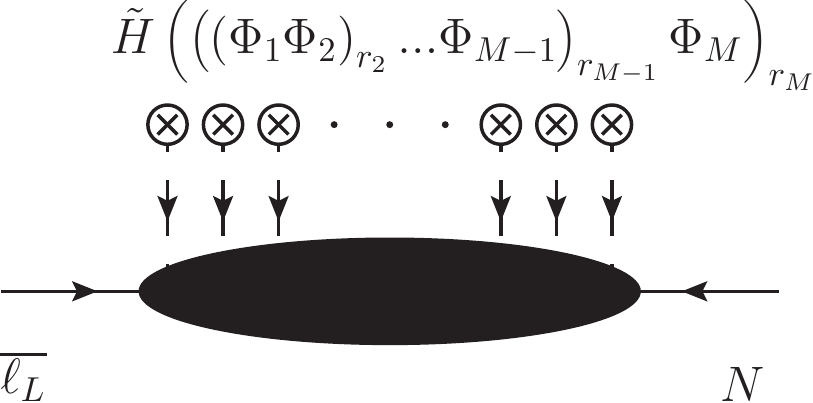}
\caption{Effective higher dimension operator from the general diagram in Figure~\ref{fig:general} after vector-like fermions are integrated out. Note that the ordering of flavons in the effective operator is indeed reversed compared with that in Figure~\ref{fig:general}.}
\label{fig:integrated}
\end{center}
\end{figure}

We assume that the mediators are very heavy $M_I\gg \langle \Phi_J \rangle$, decouple from the theory, and result in higher-dimensional operators 
\begin{eqnarray}
\frac{\lambda}{\Lambda^M}\left( \Phi_1\Phi_2\cdots \Phi_M\overline{\ell_L} \right) \tilde{H} N,
\label{eq:key_initial}
\end{eqnarray}
where the brackets denote the flavor symmetry contraction with $\overline{\ell_L}$. This is distinct from models using two flavons as $(\Phi_1 \overline{\ell_L}) (\Phi_2 N)$.
To find an explicit expression for Eq.~\eqref{eq:key_initial} in terms of the UV-complete theory, we start by assuming $N\sim \mathbf{1}$ and perform the following procedure: 
\begin{itemize}
\item Step 1, consider the terms
\begin{eqnarray}
y_2 \Phi_2 \overline{F_2} F_1 + y_1 \Phi_1 \overline{F_1} N + M_1 (\overline{F_1} F_1)_\mathbf{1} 
\end{eqnarray} 
and assume that $F_1$ decouples. Since $\Phi_2 \overline{F_2} F_1= \Big( (\Phi_2 \overline{F_2} )_{\mathbf{r}_1} F_1 \Big)_\mathbf{1}$, after $F_1$ decouples, we get in the Lagrangian the effective terms
\begin{eqnarray}
-\frac{y_1 y_2}{M_1} \Big( ( \Phi_2 \overline{F_2} )_{\mathbf{r}_1} \Phi_1 \Big)_\mathbf{1} N =
-\frac{y_1 y_2}{M_1} \Big( ( \Phi_1 \Phi_2 )_{\mathbf{r}_2} \overline{F_2}   \Big)_\mathbf{1} N \,.
\label{eq:decouple_F1}
\end{eqnarray}
\item Step 2, using Eq.\ \eqref{eq:decouple_F1} and 
\begin{eqnarray} 
y_3 \Phi_3 \overline{F_3} F_2 + M_2 (\overline{F_2} F_2)_\mathbf{1} = y_3 \Big(( \Phi_3  \overline{F_3} )_{\mathbf{r}_2} F_2 \Big)_\mathbf{1} + M_2 (\overline{F_2} F_2)_\mathbf{1} \,,
\end{eqnarray}
we assume that $F_2$ decouples. The resulting effective couplings between $\ell_L$, $\Phi_1$, $\Phi_2$ and $F_3$ can be written as
\begin{eqnarray}
(-1)^2 \frac{y_1 y_2 y_3}{i^2 M_1 M_2} \Big( ( \Phi_1 \Phi_2 )_{\mathbf{r}_2} ( \Phi_3  \overline{F_3} )_{\mathbf{r}_2} \Big)_\mathbf{1} N = 
(-1)^2 \frac{y_1 y_2 y_3}{i^2 M_1 M_2} \left( ( \Phi_1 \Phi_2 )_{\mathbf{r}_2} \Phi_3 \Big)_{\mathbf{r}_3} \overline{F_3} \right)_\mathbf{1} N
\end{eqnarray}
\item Following the same procedure, Steps 3, \dots, $M-2$. 
\item Step $M-1$, consider the decoupling of $F_{M-1}$, after which we have
\begin{eqnarray}
(-1)^{M-1} \frac{y_1 y_2 \cdots y_{M-1} y_{M}}{ M_1 M_2 \cdots M_{M-1}} \left( \left( \Big( ( \Phi_1 \Phi_2)_{\mathbf{r}_2} \cdots \Phi_{M-1} \Big)_{\mathbf{r}_{M-1}} \Phi_{M} \right)_{\mathbf{r}_M} \overline{F_M} \right)_\mathbf{1} N \,.
\end{eqnarray}
\item Step $M$, consider the decoupling of $F_{M}$, we obtain
\begin{eqnarray}
(-1)^{M-1} \frac{y y_1 y_2 \cdots y_{M-1} y_{M}}{ M_1 M_2 \cdots M_{M-1} M_{M}} \left( \left( \Big( ( \Phi_1 \Phi_2)_{\mathbf{r}_2} \cdots \Phi_{M-1} \Big)_{\mathbf{r}_{M-1}} \Phi_{M} \right)_{\mathbf{r}_M} \overline{\ell_L} \right)_\mathbf{1} \tilde{H} N \,.
\label{eq:decouple_FM}
\end{eqnarray}
\end{itemize}
Eq.\ \eqref{eq:decouple_FM} is simply written as $(Y_N \overline{\ell_{L}})_{\mathbf{1}} \tilde{H} N $
where
\begin{eqnarray}
Y_N = \frac{\lambda}{\Lambda^M} \left( \Big( ( \Phi_1 \Phi_2)_{\mathbf{r}_2} \Phi_{3} \Big)_{\mathbf{r}_{3}} \cdots \Phi_{M} \right)_{\mathbf{r}_M} \,.
\label{eq:key}
\end{eqnarray}
The above procedure is summarised in Figures~\ref{fig:general} and \ref{fig:integrated}. From the Lagrangian in Eq.~\eqref{eq:UV}, one can draw a diagram with vector-like fermionic mediators $F_I$ (for $I=1,2,...,M$) in Figures~\ref{fig:general}. After all vector-like fermions are integrated out, only an effective dimension-$(\text{dim}(M)+4)$ operator remains, which is shown in Figure~\ref{fig:integrated}. Note that the ordering of flavons in the effective operator will be reversed with respect to their ordering in the corresponding diagram, as can be seen figures \ref{fig:general} and \ref{fig:integrated} (from left to right, $\Phi_M$, $\Phi_{M-1}$ ... , $\Phi_2$, $\Phi_1$ in Figure~\ref{fig:general} and $\Phi_1$, $\Phi_2$, ... $\Phi_{M-1}$, $\Phi_M$ in Figure~\ref{fig:integrated}).

If $N\sim \mathbf{1'}$, we follow a similar procedure, and arrive at $(Y_N \overline{\ell_{L}})_{\mathbf{1'}} \tilde{H} N $
with 
\begin{eqnarray}
Y'_N = \frac{\lambda}{\Lambda^M} \left( \Big( ( \Phi_1 \Phi_2)_{\mathbf{r}'_2} \Phi_{3} \Big)_{\mathbf{r}'_{3}} \cdots \Phi_{M} \right)_{\mathbf{r}'_M} \,,
\label{eq:key2}
\end{eqnarray}
where $\mathbf{r'}=\mathbf{1'},\mathbf{1},\mathbf{2},\mathbf{3'},\mathbf{3}$ for $\mathbf{r}=\mathbf{1},\mathbf{1'},\mathbf{2},\mathbf{3},\mathbf{3'}$, respectively. 
For different right-handed neutrinos $N_i$, the choice of $\Phi_I$ and $F_I$ may be different. After including the index $i$, we finally arrive at
\begin{eqnarray}
 (Y_i \overline{\ell_{L}})_{\mathbf{1}} \tilde{H} N_i \quad \text{or} \quad 
(Y'_i \overline{\ell_{L}})_{\mathbf{1'}} \tilde{H} N_i
\end{eqnarray}
with
\begin{eqnarray}
Y_{i} = \frac{\lambda_{i}}{\Lambda^M} \left( \Big( ( \Phi_{1i} \Phi_{2i})_{\mathbf{r}_{2i}} \Phi_{3i} \Big)_{\mathbf{r}_{3i}} \cdots \Phi_{Mi} \right)_{\mathbf{3}} \,,\nonumber\\
Y'_{i} = \frac{\lambda_{i}}{\Lambda^M} \left( \Big( ( \Phi_{1i} \Phi_{2i})_{\mathbf{r}'_{2i}} \Phi_{3i} \Big)_{\mathbf{r}'_{3i}} \cdots \Phi_{Mi} \right)_{\mathbf{3'}} \,.
\label{eq:key_N}
\end{eqnarray}
where $\mathbf{r}_{Ii}$ is the representation of $F_{Ii}$, $\mathbf{r}_{Mi}$ and $\mathbf{r}'_{Mi}$ have been replaced by $\mathbf{3}$ and $\mathbf{3'}$, respectively. 
For charged leptons, assuming $l_{\alpha R} \sim \mathbf{1}$ or $\mathbf{1'}$, for $l_{\alpha R} = e_R,\mu_R,\tau_R$, we arrive at a similar structure
\begin{eqnarray}
(Y_{\alpha} \overline{\ell_{L}})_{\mathbf{1}} \tilde{H} l_{\alpha R} \quad \text{or} \quad
(Y'_{\alpha} \overline{\ell_{L}})_{\mathbf{1'}} \tilde{H} l_{\alpha R}
\end{eqnarray}
where
\begin{eqnarray}
Y_\alpha = \frac{\lambda_\alpha}{\Lambda^M} \left( \Big( ( \Phi_{1\alpha} \Phi_{2\alpha} )_{\mathbf{r}_{2\alpha}} \Phi_{3\alpha} \Big)_{\mathbf{r}_{3\alpha}} \cdots \Phi_{M\alpha} \right)_{\mathbf{3}} \,,\nonumber\\
Y'_\alpha = \frac{\lambda_\alpha}{\Lambda^M} \left( \Big( ( \Phi_{1\alpha} \Phi_{2\alpha} )_{\mathbf{r}'_{2\alpha}} \Phi_{3\alpha} \Big)_{\mathbf{r}'_{3\alpha}} \cdots \Phi_{M\alpha} \right)_{\mathbf{3'}} \,.
\label{eq:key_l} 
\end{eqnarray}
A special feature in this approach is that all these Yukawa couplings have similar flavour contractions such as $\Big(\big((\Phi_1 \Phi_2)_{\mathbf{r}_2} \Phi_3\big)_{\mathbf{r}_3} \Phi_4\Big)_{\mathbf{r}_4} \cdots$ after the fermion messengers have been integrated out. Other flavon contractions such as $\big((\Phi_1 \Phi_2)_{\mathbf{r}_2} (\Phi_3 \Phi_4)_{\mathbf{r}_3}\big)_{\mathbf{r}_4}$ are not allowed.
With the help of Eqs.\ \eqref{eq:key_N} and \eqref{eq:key_l}, we can  combine different VEVs in Eq.\ \eqref{eq:flavon_vevs} to derive different structures of $Y_i$, $Y_i'$ and $Y_\alpha$, $Y_\alpha'$, and thus obtain different structures for the lepton mass matrices $M_l$ and $M_D$. 

\subsection{Flavour structures from effective alignments \label{sec:eff_dir}}

We continue considering the flavour symmetry $S_4 \times U(1)_F$ and introduce four $S_4$ triplet flavons $\Phi_l \sim \mathbf{3}$, $\Phi_{l'}\sim \mathbf{3'}$, $\Phi_\nu\sim \mathbf{3}$ and $\Phi_{\nu'}\sim \mathbf{3'}$. They are complex and have different charges under the $U(1)_F$ symmetry. Following the conclusion in Section \ref{sec:VEV}, we can obtain the following VEVs by assuming different relations of coefficients in the potentials 
\bq 
\langle \Phi_\nu \rangle = \begin{pmatrix}1 \\ 1 \\ 1 \end{pmatrix} \frac{v_\text{I}}{\sqrt{3}}\,,~
\langle \Phi_{\nu'} \rangle = \begin{pmatrix} \frac{1}{\sqrt{3}}  \\ \frac{1}{3+\sqrt{3}}  \\ \frac{-1}{3-\sqrt{3}}  \end{pmatrix} v_\text{II}\,,~
\langle \Phi_l \rangle = \begin{pmatrix}1 \\ 0 \\ 0 \end{pmatrix} v_{\text{III}}\,,~
\langle \Phi_{l'} \rangle = \begin{pmatrix} 0 \\ 0 \\ 1 \end{pmatrix} v_{\text{III}'}\,,
\label{eq:VEVs1}
\nq 
where these directions are presented already in the second basis. $\langle \Phi_\nu \rangle$ is orthogonal with $\langle \Phi_{\nu'} \rangle$ and $\langle \Phi_l \rangle$ is orthogonal with $\langle \Phi_{l'} \rangle$. As all the VEVs are real in this basis, their CP conjugates can be directly written out by exchanging their second and third components. It is obvious that $\langle \tilde{\Phi}_\nu \rangle$ and $\langle \tilde{\Phi}_l \rangle$ take exactly the same form $\langle \Phi_\nu \rangle$ and $\langle \Phi_l \rangle$, respectively, while the second and third components of the VEVs $\langle \tilde{\Phi}_{\nu'} \rangle$ and $\langle \tilde{\Phi}_{l'} \rangle$ are exchanged from $\langle \Phi_{\nu'} \rangle$ and $\langle \Phi_{l'} \rangle$, respectively. We note that cross-couplings between different flavons exist in general, and that these flavon cross couplings may modify the VEV directions \cite{Pascoli:2016eld}. Here, we assume these couplings are not allowed and thus do not consider their influence on the VEVs. A general discussions of how the cross-couplings modify these VEV directions, and how to forbid these couplings from a top-down approach is included in Appendix \ref{sec:cross_coupling}. 

The Kronecker products of these VEVs produce further directions, which is precisely what we refer to as EAs, as they will appear in effective terms where flavons appear $M$ times (as in Section \ref{sec:UV_eff}). In Eq.\ (\ref{eq:VEVs2}), we list all possible EAs arising from the Kronecker products of up to three of the VEVs in Eq.\ (\ref{eq:VEVs1}) up to permutations of the entries of the EA appearing as the product. Note that we are not considering here all representatives of the classes that had been listed in Eq.\ (\ref{eq:flavon_vevs}), but just the first representative for each class.
\begin{eqnarray}
M=2: &&
\langle (\Phi_{\nu}{\Phi}_{l})_\mathbf{3} \rangle \propto \begin{pmatrix} 2 \\ -1 \\ -1 \end{pmatrix} \,,~
\langle ({\Phi}_{\nu}{\Phi}_{l})_\mathbf{3'} \rangle \propto \begin{pmatrix} 0 \\ 1 \\ -1 \end{pmatrix} \,,~
\nonumber\\
&&
\langle (\Phi_{\nu'}{\Phi}_{l})_\mathbf{3} \rangle \propto \begin{pmatrix} 0 \\ z \\ -z' \end{pmatrix} \,, ~
\langle (\Phi_{\nu'}{\Phi}_{l})_\mathbf{3'} \rangle \propto \begin{pmatrix} 2 \\ -z \\ -z' \end{pmatrix} \,,~\nonumber\\
&&
\langle (\Phi_{\nu'}\Phi_{l'})_\mathbf{3} \rangle \propto \begin{pmatrix} -z \\ 2z' \\ -1 \end{pmatrix} \,,~
\langle (\Phi_{\nu'}\Phi_{l'})_\mathbf{3'} \rangle \propto \begin{pmatrix} -z \\ 0 \\ 1 \end{pmatrix} \,,~\nonumber\\
&&
\langle (\Phi_{\nu'}\tilde{\Phi}_{l'})_\mathbf{3} \rangle \propto \begin{pmatrix} -z' \\ -1 \\ 2z \end{pmatrix} \,,~
\langle (\Phi_{\nu'}\tilde{\Phi}_{l'})_\mathbf{3'} \rangle \propto \begin{pmatrix} -z' \\ 1 \\ 0 \end{pmatrix} \,,~\nonumber\\
M=3: &&
\Big\langle \Big( (\Phi_{\nu}\Phi_{l})_\mathbf{3}\Phi_l \Big)_\mathbf{3} \Big\rangle \propto \begin{pmatrix} 4 \\ 1 \\ 1 \end{pmatrix}  \,,~\nonumber\\
&&
\Big\langle \Big( (\Phi_{\nu}\Phi_{l})_\mathbf{3}\Phi_{l'} \Big)_\mathbf{3} \Big\rangle \propto \begin{pmatrix} 1 \\ 0 \\ 2 \end{pmatrix} \,,~
\Big\langle \Big( (\Phi_{\nu}\Phi_{l})_\mathbf{3}\Phi_{l'} \Big)_\mathbf{3'} \Big\rangle \propto \begin{pmatrix} -1 \\ 2 \\ 2 \end{pmatrix}  \,,~\nonumber\\
&&
\Big\langle \Big( (\Phi_{\nu}\Phi_{l})_\mathbf{3'}\Phi_l \Big)_\mathbf{3} \Big\rangle \propto \begin{pmatrix} 0 \\ 1 \\ 1 \end{pmatrix}  \,,~
\Big\langle \Big( (\Phi_{\nu'}\Phi_{l})_\mathbf{3}\Phi_{l'} \Big)_\mathbf{3'} \Big\rangle \propto \begin{pmatrix} z \\ 2z' \\ 0 \end{pmatrix}  \,,~\nonumber\\
&&
\Big\langle \Big( (\Phi_{\nu'}\Phi_{l})_\mathbf{3}\tilde{\Phi}_{l'} \Big)_\mathbf{3'} \Big\rangle \propto \begin{pmatrix} z' \\ 0 \\ 2z \end{pmatrix}  \,,~
\Big\langle \Big( (\Phi_{\nu'}\Phi_{l})_\mathbf{3'}\Phi_{l} \Big)_\mathbf{3'} \Big\rangle \propto \begin{pmatrix} 4 \\ z \\ z' \end{pmatrix}  \,,~\nonumber\\
&&
\Big\langle \Big( (\Phi_{\nu'}\Phi_{l})_\mathbf{3'}\Phi_{l'} \Big)_\mathbf{3} \Big\rangle \propto \begin{pmatrix} -z \\ 2z' \\ 2 \end{pmatrix}  \,,~
\Big\langle \Big( (\Phi_{\nu'}\Phi_{l})_\mathbf{3'}\Phi_{l'} \Big)_\mathbf{3'} \Big\rangle \propto \begin{pmatrix} z \\ 0 \\ 2 \end{pmatrix}  \,,~\nonumber\\
&&
\Big\langle \Big( (\Phi_{\nu'}\Phi_{l})_\mathbf{3'}\tilde{\Phi}_{l'} \Big)_\mathbf{3} \Big\rangle \propto \begin{pmatrix} -z' \\ 2 \\ 2z \end{pmatrix}  \,,~
\Big\langle \Big( (\Phi_{\nu'}\Phi_{l})_\mathbf{3'}\tilde{\Phi}_{l'} \Big)_\mathbf{3'} \Big\rangle \propto \begin{pmatrix} z' \\ 2 \\ 0 \end{pmatrix}  \,,~\nonumber\\
&&
\Big\langle \Big( (\Phi_{\nu'}\Phi_{l'})_\mathbf{3'}\Phi_{l} \Big)_\mathbf{3} \Big\rangle \propto \begin{pmatrix} 2z \\ 2z' \\ -1 \end{pmatrix}  \,,~
\Big\langle \Big( (\Phi_{\nu'}\Phi_{l'})_\mathbf{3'}\Phi_{l} \Big)_\mathbf{3'} \Big\rangle \propto \begin{pmatrix} 0 \\ 2z' \\ 1 \end{pmatrix}  \,,~\nonumber\\
&&
\Big\langle \Big( (\Phi_{\nu'}\Phi_{l'})_\mathbf{3'}\Phi_{\nu'} \Big)_\mathbf{3'} \Big\rangle \propto \begin{pmatrix} 2z' \\ \sqrt{3} \\ -1 \end{pmatrix}  \,,~
\Big\langle \Big( (\Phi_{\nu'}\Phi_{l'})_\mathbf{3'}\Phi_{\nu'} \Big)_\mathbf{3'} \Big\rangle \propto \begin{pmatrix} -2z' \\ -2-\sqrt{3} \\ 1 \end{pmatrix}  \,,~\nonumber\\
&&
\Big\langle \Big( (\Phi_{\nu'}\Phi_{l'})_\mathbf{3'}\tilde{\Phi}_{\nu'} \Big)_\mathbf{3'} \Big\rangle \propto \begin{pmatrix} -2z \\ 1 \\ 4+\sqrt{3} \end{pmatrix}  \,,~
\Big\langle \Big( (\Phi_{\nu'}\Phi_{l'})_\mathbf{3'}\tilde{\Phi}_{l'} \Big)_\mathbf{3'} \Big\rangle \propto \begin{pmatrix} 1 \\ z \\ 4z' \end{pmatrix}  \,,~\nonumber\\
&&
\Big\langle \Big( (\Phi_{\nu'}\tilde{\Phi}_{l'})_\mathbf{3'}\Phi_{l} \Big)_\mathbf{3'} \Big\rangle \propto \begin{pmatrix} 0 \\ 1 \\ 2z \end{pmatrix}  \,,~
\Big\langle \Big( (\Phi_{\nu'}\tilde{\Phi}_{l'})_\mathbf{3'}\Phi_{l'} \Big)_\mathbf{3'} \Big\rangle \propto \begin{pmatrix} 1 \\ 4z \\ z' \end{pmatrix}  \,,~\nonumber\\
&&
\Big\langle \Big( (\Phi_{\nu'}\tilde{\Phi}_{l'})_\mathbf{3'}\Phi_{\nu'} \Big)_\mathbf{3'} \Big\rangle \propto \begin{pmatrix} 2z \\ 1 \\ \sqrt{3} \end{pmatrix}  \,,~
\Big\langle \Big( (\Phi_{\nu'}\tilde{\Phi}_{l'})_\mathbf{3'}\tilde{\Phi}_{\nu'} \Big)_\mathbf{3'} \Big\rangle \propto \begin{pmatrix} -2z' \\ 4-\sqrt{3} \\ 1 \end{pmatrix}  \,,~
\label{eq:VEVs2} 
\end{eqnarray}
where $z=1/(1+\sqrt{3})$ and $z'=1/(1-\sqrt{3})$. In any above Kronecker products, replacing all flavons by their conjugates corresponds to exchanging the second and third entries (as in Eq.(\ref{eq:triplet_swap})) of the respective EA that results from those flavons. Since $\langle \tilde{\Phi}_\nu \rangle$ and $\langle \tilde{\Phi}_l \rangle$ are identical to $\langle \Phi_\nu \rangle$ and $\langle \Phi_l \rangle$, respectively, replacing $\Phi_\nu$ with $\tilde{\Phi}_\nu$ or $\Phi_l$ with $\tilde{\Phi}_l$ will not modify any VEV directions in Eq.~\eqref{eq:VEVs2}. 
Any other Kronecker product for $M \leqslant 3$ either vanishes, or always takes one of the directions shown in Eqs.\ \eqref{eq:VEVs1} and \eqref{eq:VEVs2}, or one related by permutations of entries. While several higher order EAs will vanish or repeat lower order directions, we note that as $M\to \infty$, one obtains infinite EAs. For example, $\left\langle \Big(\cdots \big((\Phi_\nu \Phi_l)_\mathbf{3} \Phi_l \big)_\mathbf{3} \cdots  \Phi_l \Big)_\mathbf{3} \right\rangle$ with $n$ $\Phi_l$ in the contraction gives the direction proportional to $\big((-2)^{n},1,1\big)^T$, and $\left\langle \Big(\cdots \big((\Phi_{\nu'} \Phi_l)_\mathbf{3'} \Phi_l \big)_\mathbf{3'} \cdots  \Phi_l \Big)_\mathbf{3'} \right\rangle$ with $n$ $\Phi_l$ in the contraction gives the direction proportional to $\big((-2)^{n},z,z'\big)^T$.

\subsection{Lepton mixing in the CSD framework \label{sec:lep_mix}} 

In the framework of CSD, we can use the flavon VEVs in Eq.\ \eqref{eq:VEVs1} and the EAs derived from them in Eq.\ \eqref{eq:VEVs2} to realise special leptonic mixing. We introduce three right-handed neutrinos, all of which are singlets of $S_4$: $N_1, N_2, N_3\sim\mathbf{1}$ or $\mathbf{1'}$. The SM fermions and the Higgs are arranged as $\ell_L=(\ell_e,\ell_\mu,\ell_\tau)^T\sim\mathbf{3}$, $e_R\sim \tau_R \sim \mathbf{1'}$ and $\mu_R\sim H \sim \mathbf{1}$ under $S_4$. Then we write the following Lagrangian terms, 
\begin{eqnarray}
-\mathcal{L}_l &=& (Y_\tau' \overline{\ell_L})_{\mathbf{1'}} H \tau_R 
+ \big( Y_\mu \overline{\ell_L}\big)_{\mathbf{1}} H \mu_R  
+(Y_e' \overline{\ell_L} )_{\mathbf{1'}} H e_R 
+ \text{h.c.} \,, \nonumber\\
-\mathcal{L}_\nu &=& \sum_{i=1,2,3}\, (Y_i^{(\prime)} \overline{\ell_L})_{\mathbf{1}^{(\prime)}} \tilde{H} N_i + \frac{1}{2} M_i\overline{N^c_i} N_i + \text{h.c.} \,, 
\label{eq:L_N}
\end{eqnarray} 
where the choice of $\mathbf{1}$ or $\mathbf{1}'$ for the combination of $Y_i^{(\prime)} \overline{\ell_L}$ depends on the representation of $N_i$. 
There are a lot of ways to arrange the charged lepton Yukawa coupling as 
\bq
Y_e\propto \begin{pmatrix} 1 \\ 0 \\ 0 \end{pmatrix}\,,\quad
Y_\mu\propto \begin{pmatrix} 0 \\ 1 \\ 0 \end{pmatrix}\,,\quad
Y_\tau\propto \begin{pmatrix} 0 \\ 0 \\ 1 \end{pmatrix}\,,\quad
\nq 
from the flavon VEVs and their EAs. For example, $Y_e \propto \langle ((\Phi_{l'} \Phi_{l'})_{\mathbf{3}} \Phi_{l'})_\mathbf{3, 3'} \rangle $, $Y_\mu \propto \langle (\Phi_{l'} \Phi_{l'})_{\mathbf{3}} \rangle$, $Y_\tau \propto \langle \Phi_{l'} \rangle$. 
From these couplings, charged leptons obtain a diagonal mass matrix, and all flavour mixing comes from the neutrino Yukawa couplings $Y_i^{(\prime)}$. The latter can in principle take any combination of three directions from Eqs.~\eqref{eq:VEVs1} and \eqref{eq:VEVs2}. We normalise $Y_i^{(\prime)}$ to $V_i$, respectively, with $V_i^\dag V_i=1$ required, therefore $V_i$ now denote the normalised (and dimensionless) directions of the VEVs. 
The final active neutrino mass matrix takes the form 
\bq
M_\nu &=& \sum_{i=1,2,3}\, \frac{v^2}{M_{N_i}} Y_i^{(\prime)} Y_i^{(\prime)T}  
= \sum \mu_i V_i V_i^T \,,
\label{eq:neutrino_mass}
\nq
where $\mu_i$ are complex mass parameters, unspecified by the flavour symmetry. The VEV combinations $V_i$ and the parameters $\mu_i$ fully determine the mixing. In the original idea of CSD models, a hierarchy of $\mu_1$, $\mu_2$ and $\mu_3$ is assumed based on a strong hierarchy of the right-handed neutrino masses, but here we do not make this assumption unless mentioned explicitly.

The use of VEV directions in CSD models with 2 only RHNs is particularly predictive as it corresponds to one $\mu_i$ being zero, and therefore one of the active neutrinos having zero mass, by the following argument.  For Normal Hierarchy (NH), $m_1=0$, as with the first column of the PMNS matrix $U_1$ holds that $U_1^\dag M_\nu =m_1 U_1^T =0$. From Eq.~\eqref{eq:neutrino_mass} with $\mu_3=0$ we get $M_\nu = \mu_1 V_1 V_1^T + \mu_2 V_2 V_2^T $. Multiplying $U_1^\dag$ from the left of $M_\nu$, we obtain $(\mu_1 U_1^\dag V_1) V_1^T + (\mu_2 U_1^\dag V_2) V_2^T=0$. If $V_1^T$ is linearly independent of $V_2^T$, we must have vanishing coefficients, i.e., $U_1^\dag V_1 = U_1^\dag V_2=0$. Thus, once we choose two VEV directions ($V_1$ and $V_2$) from Eqs.~\eqref{eq:VEVs1} and \eqref{eq:VEVs2}, we can determine the first column of the PMNS matrix $U_1$ directly. Based on the relation between $V_1$ and $V_2$, we emphasise the following two cases: 
\begin{itemize}
\item $V_1$ is orthogonal to $V_2$, $V_1^\dag V_2=0$: It directly leads to the PMNS mixing matrix 
\bq
U_\text{PMNS}=(U_1, V_1, V_2) \;\;\text{or}\;\; (U_1, V_2, V_1) \,,
\nq 
where we don't show the free Majorana phases for simplicity.
In this case, the mixing matrix is a constant mixing pattern with three columns independent of the neutrino masses (also referred to as mass-independent mixing schemes or form-diagonalizable schemes \cite{deMedeirosVarzielas:2011tp}). The neutrino mass eigenvalues are $\{m_1, m_2,  m_3\}=\{0, |\mu_1|, |\mu_2|\}$ or $\{0, |\mu_2|, |\mu_1|\}$, respectively. It is very easy to realise some constant mixing pattern. For example, the tri-bimaximal (TBM) mixing
\bq
U_{\text{TBM}} = \left(
\begin{array}{ccc}
 \frac{2}{\sqrt{6}} & \frac{1}{\sqrt{3}} & 0 \\
 \frac{-1}{\sqrt{6}} & \frac{1}{\sqrt{3}} & \frac{1}{\sqrt{2}} \\
 \frac{-1}{\sqrt{6}} & \frac{1}{\sqrt{3}}  & \frac{-1}{\sqrt{2}} \\
\end{array}
\right)\,.
\nq
is realised by picking from the available directions
\bq
V_1 \propto \langle \Phi_\nu \rangle \,, \quad 
V_2 \propto \langle (\Phi_{l}\Phi_{\nu})_\mathbf{3'} \rangle \,.
\nq
Another constant mixing which can be similarly realised is the Toorop-Feruglio-Hagedorn (TFH) mixing \cite{Toorop:2011jn}
\bq
U_\text{TFH} &=& \left(
\begin{array}{ccc}
\frac{-1}{3-\sqrt{3}} & \frac{1}{\sqrt{3}} & \frac{1}{3+\sqrt{3}} \\ 
\frac{1}{\sqrt{3}} & \frac{1}{\sqrt{3}} & \frac{1}{\sqrt{3}} \\ 
\frac{1}{3+\sqrt{3}} & \frac{1}{\sqrt{3}} & \frac{-1}{3-\sqrt{3}}
\end{array}
\label{eq:U_TFH}
\right)
\nq
This pattern predicts non-zero $\theta_{13}$ at leading order. 
It is obtained by choosing
\bq
V_1 \propto \langle \Phi_\nu \rangle \,,\quad
V_2 \propto \langle \Phi_{\nu'} \rangle \,.
\nq
However, these patterns have been experimentally excluded by current data. 
To be consistent with current data, small corrections, $\delta V_2$ or $\delta V_3$ would need to be included (see e.g.\ \cite{Sierra:2013ypa, Sierra:2014hea}). 

\item $V_1$ is not orthogonal to $V_2$: The mixing matrix can be parametrised by a rotation between the 2nd and 3rd columns,
\bq
U_\text{PMNS}=(U_1, V_1, V_2) \begin{pmatrix} 1 & 0 & 0 \\ 0 & \cos\theta & \sin\theta e^{-i \gamma} \\ 0 & -\sin\theta e^{i \gamma} & \cos\theta \end{pmatrix} \,,
\nq 
where $\theta$ and $\gamma$ are functions of $\mu_1/\mu_2$, i.e.\ the neutrino mass ratio. This is a partially constant mixing, where the first column is independent of neutrino masses. A common example is TM1 mixing \cite{Xing:2006ms, Lam:2006wm, Albright:2008rp, Albright:2010ap}, which arises when $V_1, V_2 \perp (2, -1, -1)^T$ and can be obtained adequately with $S_4$ \cite{Varzielas:2012pa} or other groups \cite{Varzielas:2015aua}. Another example is the CSD2 model with two RHNs. In this model,
\bq
\label{eq:unperturbed}
V_1 \propto \Big\langle \Big( (\Phi_{l}\Phi_{\nu})_\mathbf{3}\tilde{\Phi}_{l'} \Big)_\mathbf{3} \Big\rangle \,, \quad
V_2 \propto \langle (\Phi_{l}\Phi_{\nu})_\mathbf{3'} \rangle \,.
\nq
With these alignments, the neutrino mass matrix is given by
\bq
M_\nu = 
\frac{\mu_2}{2} \left(
\begin{array}{ccc}
0 &0 & 0 \\ 
0 &1 & -1 \\ 
0 &-1 & 1
\end{array}
\right)
+  \frac{\mu_1}{5} 
 \left(
\begin{array}{ccc}
1 &2 & 0 \\ 
2 &4 & 0 \\ 
0 &0 & 0
\end{array}
\right) \,.
\nq
For further discussion of this scenario, please see \cite{Antusch:2011ic, Bjorkeroth:2014vha}. 
\end{itemize}
On the other hand, for the case of Inverted Hierarchy (IH) with two RHNs, since $m_3=0$ holds as the with the third column of the PMNS matrix $U_3$ follows that $U_3^\dag M_\nu =m_3 U_3^T =0$, one obtains $U_3^\dag V_1 = U_3^\dag V_2=0$. Thus, similarly to the NH case, once we pick up two VEV directions ($V_1$ and $V_2$) from Eqs.~\eqref{eq:VEVs1} and \eqref{eq:VEVs2}, we can determine $U_3$ directly. If in addition $V_2^\dag V_3=0$, then the PMNS matrix is given by $U_{\text{PMNS}} = (V_1, V_2, U_3)$ or $(V_2, V_1, U_3)$, ignoring Majorana phases. If $V_1^\dag V_2\neq0$, only the third column of the PMNS is mass-independent. As for the first entry of $U_3$, $|U_{13}|=\sin\theta_{13}$, one has to carefully choose $V_1$ and $V_2$ in this scenario. 
We already showed an example leading to the TFH mixing in Eq.(\ref{eq:U_TFH}), by selecting $V_1 = \Phi_{\nu'}$ and $V_2 = \Phi_\nu$.
In the sense that it produces a good value for the reactor angle (but not of the other mixing angles), another interesting example of $U_3$ being defined by the choice of $V_1$,$V_2$ with the EAs obtained in $S_4$ is taking $V_1$, $V_2$ as $((-2)^{N+1},1,1)^T$ and  $((-2)^N,z,z')^T$. This choice determines $U_3 \propto (1, (-2)^N, (-2)^N)^T$, corresponding to a reactor angle of $\sin^2\theta_{13} = 1/(1+2^{2N+1})$. For $N=2$, $\sin^2\theta_{13} = 1/33$, close to the central experimental value.

In the three RHNs case, the mixing matrix and mass eigenvalues are in general functions of just $V_{1,2,3}$ and $\mu_{1,2,3}$, and mixing parameters and masses are correlated with each other.
One special case is that $V_1$, $V_2$, $V_3$ are orthogonal with each other, $V_i^\dag V_j = \delta_{ij}$. In that case, up to the Majorana phases which we do not show, the mixing matrix is directly given by $U_{\text{PMNS}} = (V_1, V_2, V_3)$ up to permutation of columns, and the mass eigenvalues are just $|\mu_1|$, $|\mu_2|$ and $|\mu_3|$. 
Another interesting case is, as originally proposed in the CSD framework \cite{King:2005bj, Antusch:2011ic, Bjorkeroth:2014vha}, that of hierarchical structures among $\mu_1$, $\mu_2$ and $\mu_3$, e.g.\ $|\mu_3| \ll |\mu_1|< |\mu_2|$. There, one may treat the lightest mass parameter as a small perturbation that produces corrections at the level of $\mu_3/\mu_1$ or  $\mu_3/\mu_2$ to the PMNS matrix derived with 2RHNs. 

\section{CSD2 models \label{sec:CSD2 models}}

We intend now to build a few specific models using EAs.
In general, without any symmetry distinguishing the flavons, all different higher order operators can contribute, such that the full effective Yukawa coupling to right-handed neutrinos becomes
\begin{equation}
 H\left[\left(l_{\bf 3}(\phi_1+\ldots+\phi_1^2+\phi_1 \phi_2+\ldots+\phi_i\phi_j\ldots\phi_n+\ldots)_{\bf 3^{(')}}\right)_{\bf 1^{(')}} N_i\right]_{\bf 1}.
\end{equation}
In front of every contribution in the sum, a coefficient has been omitted. In order to increase the predictivity of the model, one should avoid having more than one term consisting of different flavon combinations in the coupling to each right-handed neutrino, as more combinations introduce additional free parameters.
We are going to achieve this by having the flavons be charged under $U(1)_F$. This still allows multiple orderings of the same flavon combinations, e.g.\ if $\phi_1 \phi_2$ is allowed by $U(1)_F$, then $\phi_2 \phi_1$ is also allowed and may correspond to a different EA. This in turn will be avoided by constructing UV complete theories that allow only one of the orderings in the needed effective operator.

The models will be built in the CSD framework. In models of this type, two directions are fixed at $V_1=(1,2,0)^T/\sqrt{5}$ and  $V_2=(0,1,-1)^T/\sqrt{2}$, thus we refer to them as CSD2 models. The only difference is the third VEV direction, $V_3=(0,0,1)^T$, $(1,1,1)^T$ and $(2,-1,-1)^T$, in Models I, II, and III, respectively. 
The active neutrino mass matrix in these models are exactly written out as
\bq
\text{Model I: }&& M_\nu = 
\hspace{.5cm}\mu_3
 \left(
\begin{array}{ccc}
0 &0 & 0 \\ 
0 &0 & 0 \\ 
0 &0 & 1
\end{array}
\right) \hspace{.5cm}
+ \frac{\mu_2}{2} \left(
\begin{array}{ccc}
0 &0 & 0 \\ 
0 &1 & -1 \\ 
0 &-1 & 1
\end{array}
\right)
+ \frac{\mu_1}{5} 
 \left(
\begin{array}{ccc}
1 &2 & 0 \\ 
2 &4 & 0 \\ 
0 &0 & 0
\end{array}
\right)
  \,;\nonumber\\
\text{Model II: }&& M_\nu = 
\hspace{.5cm}\frac{\mu_3}{3} 
 \left(
\begin{array}{ccc}
1 &1 & 1 \\ 
1 &1 & 1 \\ 
1 &1 & 1
\end{array}
\right) \hspace{.4cm}
+ \frac{\mu_2}{2} \left(
\begin{array}{ccc}
0 &0 & 0 \\ 
0 &1 & -1 \\ 
0 &-1 & 1
\end{array}
\right)
+ \frac{\mu_1}{5} 
 \left(
\begin{array}{ccc}
1 &2 & 0 \\ 
2 &4 & 0 \\ 
0 &0 & 0
\end{array}
\right)
  \,;\nonumber\\
\text{Model III: }&& M_\nu = 
\frac{\mu_3}{6} 
 \left(
\begin{array}{ccc}
4 &-2 & -2 \\ 
-2 &1 & 1 \\ 
-2 &1 & 1
\end{array}
\right) 
+ \frac{\mu_2}{2} \left(
\begin{array}{ccc}
0 &0 & 0 \\ 
0 &1 & -1 \\ 
0 &-1 & 1
\end{array}
\right)
+ \frac{\mu_1}{5} 
 \left(
\begin{array}{ccc}
1 &2 & 0 \\ 
2 &4 & 0 \\ 
0 &0 & 0
\end{array}
\right)
 \,.
 \label{mass_matrices_of_models_123}
\nq
Model I can be understood as the original CSD2 model \cite{Antusch:2011ic}, but with three right-handed neutrinos, which allows $\mu_3 \neq 0$. Models II and III preserve TM1 mixing. Note in Model II that the three VEV directions are not linearly independent of each other, and thus the lightest neutrino mass in Model II (corresponding to the eigenvector $(2,-1,-1)$) vanishes. A phenomenological analysis corresponding to these mass matrices is presented in Section \ref{sec:pheno}.

\subsection{Example models}

In this section we employ our technique of EAs, obtained through combining 2 or more VEVs, in order to build a model with directions similar to those of CSD models.

Recall that the VEVs or EAs to be used in the models of Eq.\ (\ref{mass_matrices_of_models_123}) arise respectively via
\begin{align}
\langle \Phi_{l'} \rangle \propto \begin{pmatrix} 0 \\ 0 \\ 1 \end{pmatrix} \,,~
\langle \tildex{\Phi}_{\nu} \rangle \propto \begin{pmatrix} 1 \\ 1 \\ 1 \end{pmatrix} \,,~
\langle (\tildex{\Phi}_{\nu}\tildex{\Phi}_{l})_\mathbf{3} \rangle \propto \begin{pmatrix} 2 \\ -1 \\ -1 \end{pmatrix} \,,\nonumber\\
\langle (\tildex{\Phi}_{\nu}\tildex{\Phi}_{l})_\mathbf{3'} \rangle \propto \begin{pmatrix} 0 \\ 1 \\ -1 \end{pmatrix} \,,~
\Big\langle \Big( (\tildex{\Phi}_{\nu}\tildex{\Phi}_{l})_\mathbf{3}\tilde{\Phi}_{l'} \Big)_\mathbf{3} \Big\rangle \propto \begin{pmatrix} 1 \\ 2 \\ 0 \end{pmatrix}  \,.~
\label{eq:CSD2x_VEVs}
\end{align}
where the tilde refers to the conjugation shown in Eq.(\ref{eq:triplet_swap}), which swaps the second and third directions, which does not affect the directions shown above.
In principle, model I should be constructed by selecting the first, fourth and fifth directions, while model II or III is obtained by replacing the first direction with the second or third direction, respectively.

At the level of model building, our method runs into an issue with respect to EAs obtained from combining 3 (or more) VEVs - more than one different direction can arise from combining those 3 VEVs in a different ordering. In this case, at the non-renormalizable level the symmetries alone can not forbid the other orderings and two of the many directions that would arise are the permutations of the $(1,2,0)$ direction:
\begin{align}
\Big\langle \Big( (\Phi_{l}\Phi_{\nu})_\mathbf{3'}\tilde{\Phi}_{l'} \Big)_\mathbf{3} \Big\rangle \propto \begin{pmatrix} 1 \\ 0 \\ 2 \end{pmatrix}  \,,~
\Big\langle \Big( (\tilde{\Phi}_{l'}\Phi_{\nu})_\mathbf{3'}\Phi_{l} \Big)_\mathbf{3} \Big\rangle \propto \begin{pmatrix} 2 \\ 1 \\ 0 \end{pmatrix}  \,,~
\end{align}
corresponding to having a $\tilde{\Phi}_l'$ (here we omit the optional conjugations of Eq.(\ref{eq:triplet_swap}) on $\Phi_\nu$ and $\Phi_l$ that do not change the directions).
In order to preserve the models' predictivities and to have them remain viable, we want to have only three of the EAs in Eq.(\ref{eq:CSD2x_VEVs}) when constructing the Dirac neutrino mass matrix. In order to do this, we construct UV complete models with fermionic messengers such that the renormalizable terms allow only the EAs wanted in that model.

\subsubsection{Model I}

For model I, we write the following terms
\bq
\mathcal{L}_{\nu} &=& \overline{\ell_L} \tilde{H} A +  \overline{A} \Phi_l B  + \overline{A} \tilde{\Phi}_{l'} C + \overline{C} \tilde{\Phi}_l D + \overline{D} \tilde{\Phi}_\nu N_1  + \overline{B} \tilde{\Phi}_\nu N_2 + \overline{A} \Phi_{l'} N_3 \\
&+& M_A \overline{A} A + M_B \overline{B} B + M_C \overline{C} C+ M_D \overline{D} D 
+ M_1 \overline{N_1^c} N_1 + M_2 \overline{N_2^c} N_2 + M_3 \overline{N_3^c} N_3 + \text{h.c.}
\nq
with messengers $A$, $B$, $C$ and $D$ having generic mass terms denoted by $M$, and where we omitted the dimensionless couplings. 

A possible solution for $U(1) \times Z_2$ charges that achieves this without accidental terms is listed in Table \ref{CSD2x_charges1}. In order to avoid Majorana mass terms for the vector-like fermions, which we denoted by $A$, $B$, $C$ and $D$ in the model, we arrange $U(1)_F$ charges for all of them. Each flavon also transforms non-trivially under $U(1)_F$ to keep their potentials even and eliminate cross-couplings between different flavon fields, cf.\ section \ref{sec:VEV}. In addition we have also the Majorana terms $\overline{N_i^c} N_i$ with masses $M_{N_i}$, and  cross-terms like $\overline{N_1^c} N_2$ are forbidden by the assignments under two $Z_2$ symmetries which we refer to as $Z_2^a$ and $Z_2^b$ charges. After the vector-like fermions are integrated out, we obtain effective higher-dimensional operator up to dimension 7
\bq
\frac{1}{M^3}\Big( (\tilde{\Phi}_{\nu}\tilde{\Phi}_{l})_\mathbf{3}\tilde{\Phi}_{l'} \Big)_\mathbf{3} \overline{\ell_L} \tilde{H} N_1\,,\quad 
\frac{1}{M^2}(\tilde{\Phi}_{\nu} {\Phi}_{l})_\mathbf{3'} \overline{\ell_L} \tilde{H} N_2\,,\quad 
\frac{1}{M}\Phi_{l'} \overline{\ell_L} \tilde{H} N_3\,.
\nq
Due to the requirement of $U(1)_F$ symmetry, all the other higher-dimensional operators contributing to neutrino Yukawa couplings appear at  dimension $\geqslant 8$. 
As described in general in Section \ref{sec:UV_eff}, after integrating out the vector-like fermions, the order of the flavons is reversed.

\begin{figure}
\begin{center}
\includegraphics[scale=1.0]{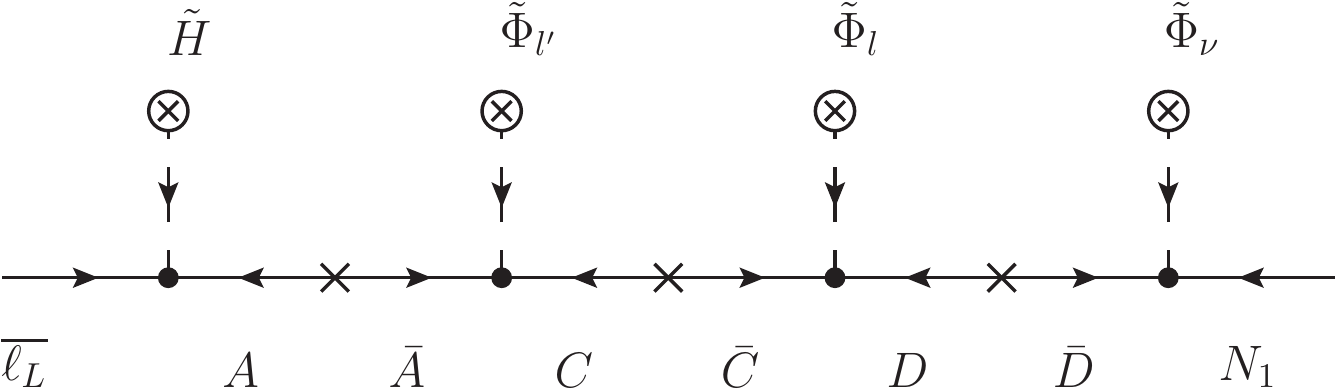}
\caption{Diagram with UV completion for $N_1$ in models I, II and III.}
\label{fig:N1}
\end{center}
\end{figure}

\begin{figure}
\begin{center}
\includegraphics[scale=1.0]{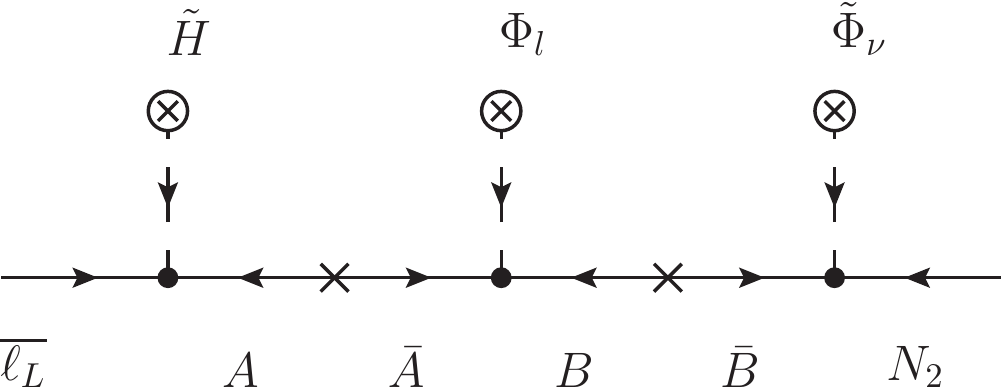}
\caption{Diagram with UV completion for $N_2$ in models I, II and III.}
\label{fig:N2}
\end{center}
\end{figure}

\begin{figure}
\begin{center}
\includegraphics[scale=1.0]{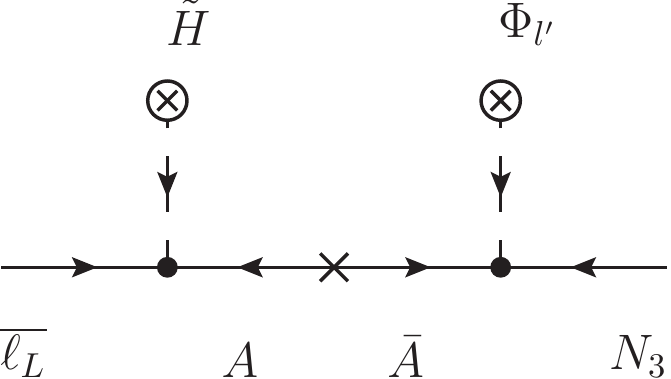}
\caption{Diagram with UV completion for $N_3$ in model I.}
\label{fig:N3_1}
\end{center}
\end{figure}

\begin{figure}
\begin{center}
\includegraphics[scale=1.0]{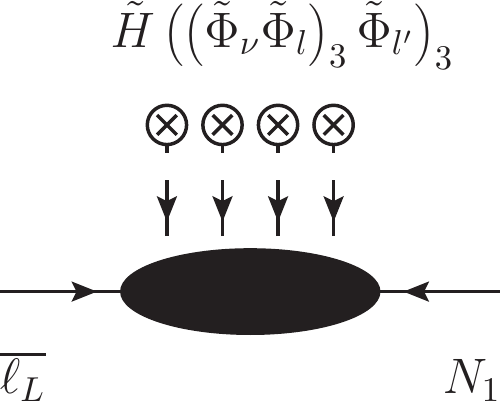}
\caption{Diagram with the effective diagram for $N_1$ in models I, II and III.}
\label{fig:N1_integrated}
\end{center}
\end{figure}

The respective diagrams are shown in Figures \ref{fig:N1}, \ref{fig:N2} and \ref{fig:N3_1}. We show also the diagram corresponding to the effective operator for $N_1$ in Figure \ref{fig:N1_integrated}, with the reversed ordering of the flavons (cf.\ Section \ref{sec:UV_eff} and Figures \ref{fig:general} and \ref{fig:integrated}).

\begin{table}[h!] \centering
    \begin{tabular}{|c|cc|ccc|ccc|cccc|}
      \hline
      Field & $\ell_L$ & $H$ & $N_1$ & $N_2$ & $N_3$ & $\Phi_{l'}$ & $\Phi_l$ & $\Phi_{\nu}$ & $A$ & $B$ & $C$ & $D$ \\ \hline
      $S_4$ & $\mathbf{3}$ & $\mathbf{1}$ & $\mathbf{1}$ & $\mathbf{1'}$ & $\mathbf{1}$ & $\mathbf{3'}$ & $\mathbf{3}$ & $\mathbf{3}$ & $\mathbf{3}$ & $\mathbf{3}$ & $\mathbf{3'}$ & $\mathbf{3}$ \\
      $U(1)_F$ & $2$ & $0$ & $0$ & $0$ & $0$ & $2$ & $-1$ & $-3$ & $2$ & $3$ & $4$ & $3$ \\
      $Z_2^a$ & $0$ & $0$ & $1$ & $1$ & $0$ & $0$ & $1$ & $0$ & $0$ & $1$ & $0$ & $1$ \\
      $Z_2^b$ & $0$ & $0$ & $1$ & $0$ & $1$ & $1$ & $0$ & $0$ & $0$ & $0$ & $1$ & $1$ \\ \hline
    \end{tabular}
    \caption{Fields and charges for Model I. \label{CSD2x_charges1}}
\end{table}

The charged lepton sector gives rise to a diagonal mass matrix through the following effective terms, using only directions obtained from powers of $\Phi_{l'}$ (see Eq.(\ref{eq:VEVs2})):
\begin{align}
\mathcal{L}_e = H \overline{\ell_L} \left( ((\Phi_{l'} \Phi_{l'})_{\mathbf{3}} \Phi_{l'})_\mathbf{3'} e_R +  (\Phi_{l'} \Phi_{l'})_{\mathbf{3}} \mu_R + \Phi_{l'} \tau_R \right)
\end{align}
where the charge assignments are that $\tau_R$ and $e_R$ are both $Z_2^b$-odd and $1'$ under $S_4$ (recall $\Phi_{l'}$ is a $3'$ so this is required to make the $S_4$ invariant contraction with $\overline{\ell_L}$). $\mu_R$ is $Z_2^b$-even and a $1$ of $S_4$. The $U(1)_F$ charges are respectively $-4$, $-2$ and $0$ for $e_R$, $\mu_R$ and $\tau_R$
\footnote{The contraction $H \overline{\ell_L} (\tilde{\Phi}_{l} \tilde{\Phi}_{l})_{\mathbf{3'}}) \tau_R$ is allowed but we omit it as the respective direction vanishes.}.
A possible UV completion for these terms requires its own set of messengers which have different $U(1)_Y$ assignments compared to the ones of the neutrino sector (due to neutrino terms featuring $\tilde{H}$ instead of $H$).

\subsubsection{Model II}

For model II, we write the following terms
\bq
\mathcal{L}_{\nu} &=& \overline{\ell_L} \tilde{H} A +  \overline{A} \Phi_l B  + \overline{A} \tilde{\Phi}_{l'} C + \overline{C} \tilde{\Phi}_l D + \overline{D} \tilde{\Phi}_\nu N_1  + \overline{B} \tilde{\Phi}_\nu N_2 + \overline{A} \Phi_{\nu} N_3 \\
&+& M \overline{A} A + M \overline{B} B + M \overline{C} C+ M \overline{D} D 
+ M_1 \overline{N_1^c} N_1 + M_2 \overline{N_2^c} N_2 + M_3 \overline{N_3^c} N_3 + \text{h.c.}
\nq
where we again generically denoted messenger masses with $M$, and omitted the dimensionless couplings.

A possible solution for $U(1) \times Z_2$ charges that achieves this without accidental terms is listed in Table \ref{CSD2x_charges}. Apart from the different charges, the model II is similar to model I, all vector-like fermions and flavons have $U(1)_F$ charges and cross-terms between the $N_i$ are forbidden by distinct $Z_2$ charges. After the vector-like fermions are integrated out, we obtain effective higher-dimensional operator up to dimension 7
\bq
\frac{1}{M^3}\Big( (\tilde{\Phi}_{\nu}\tilde{\Phi}_{l})_\mathbf{3}\tilde{\Phi}_{l'} \Big)_\mathbf{3} \overline{\ell_L} \tilde{H} N_1\,,\quad 
\frac{1}{M^2}(\tilde{\Phi}_{\nu} {\Phi}_{l})_\mathbf{3'} \overline{\ell_L} \tilde{H} N_2\,,\quad 
\frac{1}{M}\Phi_{\nu} \overline{\ell_L} \tilde{H} N_3\,,
\nq

\begin{figure}
\begin{center}
\includegraphics[scale=1.0]{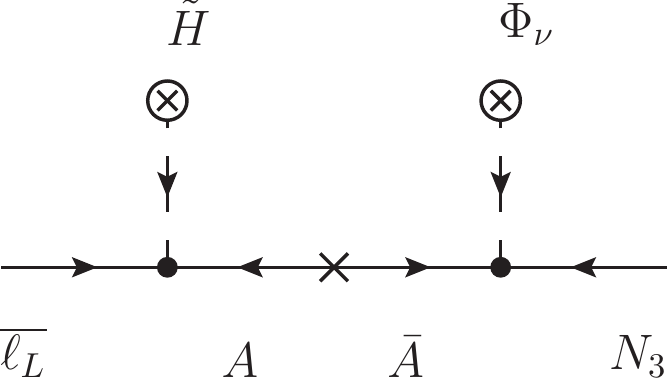}
\caption{Diagram with UV completion for $N_3$ in model II.}
\label{fig:N3_2}
\end{center}
\end{figure}

Despite the different charges, the model shares two diagrams with Model I, and thus the respective diagrams are shown in Figures \ref{fig:N1}, \ref{fig:N2} and \ref{fig:N3_2}.

\begin{table}[h!] \centering
    \begin{tabular}{|c|cc|ccc|ccc|cccc|}
      \hline
      Field & $\ell_L$ & $H$ & $N_1$ & $N_2$ & $N_3$ & $\Phi_{l'}$ & $\Phi_l$ & $\Phi_{\nu}$ & $A$ & $B$ & $C$ & $D$ \\ \hline
      $S_4$ & $\mathbf{3}$ & $\mathbf{1}$ & $\mathbf{1}$ & $\mathbf{1'}$ & $\mathbf{1}$ & $\mathbf{3'}$ & $\mathbf{3}$ & $\mathbf{3}$ & $\mathbf{3}$ & $\mathbf{3}$ & $\mathbf{3'}$ & $\mathbf{3}$ \\
      $U(1)_F$ & $1$ & $0$ & $0$ & $0$ & $0$ & $-4$ & $2$ & $1$ & $1$ & $-1$ & $-3$ & $-1$ \\
      $Z_2^a$ & $0$ & $0$ & $1$ & $1$ & $0$ & $0$ & $1$ & $0$ & $0$ & $1$ & $0$ & $1$ \\
      $Z_2^b$ & $0$ & $0$ & $1$ & $0$ & $0$ & $1$ & $0$ & $0$ & $0$ & $0$ & $1$ & $1$ \\ \hline
    \end{tabular}
    \caption{Fields and charges for Model II. \label{CSD2x_charges}}
\end{table}

Similarly to Model I, one can arrange the charged lepton sector to give rise to a diagonal mass matrix through effective terms using directions obtained from powers of $\Phi_{l'}$, although the charge assignments need to be altered. The $Z_2$ and $S_4$ assignments remain, with $\tau_R$ and $e_R$ both $Z_2^b$-odd and $1'$ under $S_4$ and $\mu_R$ being $Z_2^b$-even and a $1$ of $S_4$. The $U(1)_F$ charges become respectively $13$, $9$ and $5$ for $e_R$, $\mu_R$ and $\tau_R$
This is only one of many possibilities of obtaining the required diagonal Yukawa terms from our set of flavon VEVs, which we choose despite the relatively high $U(1)_F$ charges because it uses the same terms as Model I, which automatically incorporate a suppression of the masses for the lighter generations (which appear only as higher-dimensional operators).

\subsubsection{Model III}

\begin{figure}
\begin{center}
\includegraphics[scale=1.0]{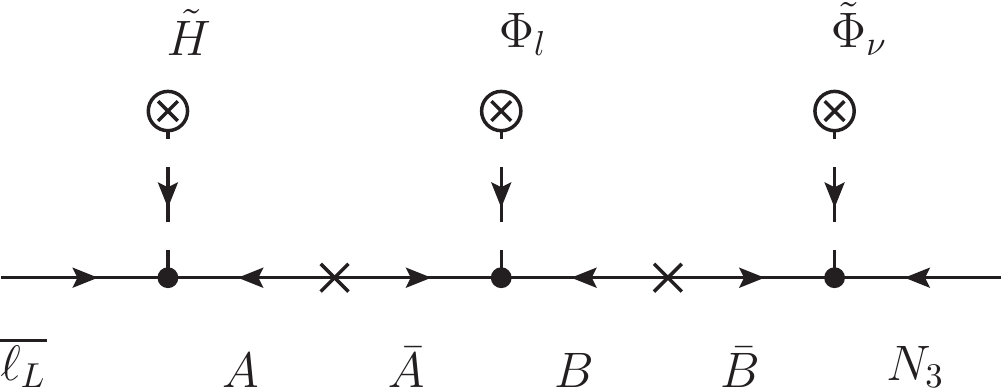}
\caption{Diagram with UV completion for $N_3$ in model III.}
\label{fig:N3_3}
\end{center}
\end{figure}

Model III is obtained by re-arranging the $Z_2^a$ charge of $N_3$ as $1$. Then, the coupling $\overline{A} \Phi_{\nu} N_3$ is forbidden, but the coupling $\overline{B} \tilde{\Phi}_\nu N_3$ is allowed. The rest of the Lagrangian is not modified. With this rearrangement, the effective higher-dimensional operator for generating the $N_3$ Yukawa structure is altered to 
\bq
\frac{1}{M^2}(\tilde{\Phi}_{\nu} {\Phi}_{l})_\mathbf{3} \overline{\ell_L} \tilde{H} N_3\,.
\nq
Model III shares two diagrams with Model I and II, with the respective diagrams shown in Figures \ref{fig:N1}, \ref{fig:N2} and \ref{fig:N3_3}.
Similarly, Model III has the same charged lepton sector as Model II above.

\subsection{Phenomenology of the CSD2 models \label{sec:pheno}}

The mixing parameters of models I, II and III depend on their respective fixed EAs and on the $\mu_i$ parameters. In this section we present results from a numerical exploration of the three models, comparing with a global fit with the following $3\sigma$ ranges for the three mixing angles and mass-square differences \cite{Esteban:2016qun} in normal hierarchy (NH),
\begin{eqnarray} 
&\theta_{12} \in [31.38^\circ,\, 35.99^\circ]\,,~ \theta_{23} \in [38.4^\circ,\, 52.8^\circ]\,,~  \theta_{13} \in [7.99^\circ,\, 8.90^\circ]\,,\nonumber\\
&\Delta m^2_{21} \in[7.03,\, 8.09]\times10^{-5}\,\text{eV}^2 \,,~ \Delta m^2_{31} \in[2.407,\, 2.643]\times10^{-3}\,\text{eV}^2.
\end{eqnarray}
The aim was to generate a sufficient number of points in the space of input parameters of each of the models such that possible correlations that are not easy to see analytically become apparent. For each those random points in parameter space, the mixing angles, masses, $\delta_\text{CP}$, and $|m_{ee}|=|(M_\nu)_{11}|$ (relevant for neutrinoless double-beta decay, cf.\ \cite{Patrignani:2016xqp}, section 14.4) had been calculated and a point would be kept if the calculated observables were within their respective three-sigma ranges from the global fit. All other points were discarded.
The input parameters of each model were expressed as $\mu_i=|\mu_i|e^{i \text{Arg}(\mu_i)}$. Furthermore, the overall scale of neutrino masses cannot be constrained by above data, which is why an overall factor $\mu_0$ was pulled out of every $\mu_i$, which for valid points was set such that the calculated $\Delta m_{31}^2$ would attain its central value from the global fit. This parametrization has the additional effect that only the ratio of mass splittings $(\Delta m_{31}^2)/(\Delta m_{21}^2)$, in which $\mu_0$ cancels, has to be compared with data.

Hence, for each model $10^7$ points were generated with parameter values $|\mu_i|\in[0,1]$ and $\text{Arg}(\mu_i)\in[0,2\pi]$. For each model, every observable was plotted against every other observable, and the most interesting of these plots are shown in this section, plotting the valid points of each model on top of each other to allow for comparison between the models. 

Our numerical results show that more points are allowed  in models II and III than in model I.

\begin{figure}
\begin{center}
\includegraphics[scale=0.4]{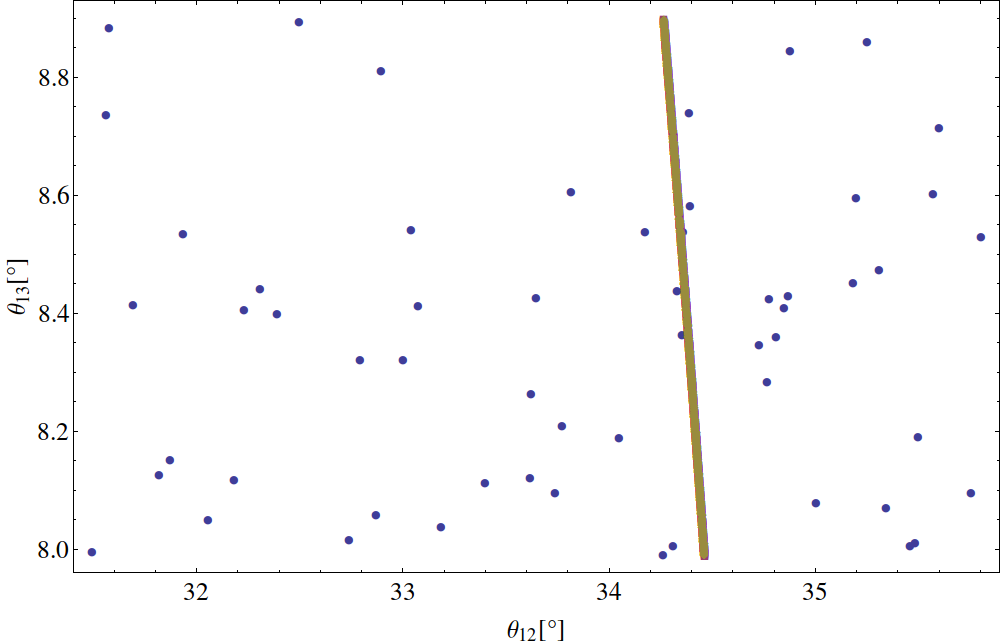}
\caption{Plot of the correlation between $\theta_{12}$ and $\theta_{13}$ in models I (blue circles), II (purple squares) and III (brown diamonds). Model II and III have the same correlation so the points overlap.}
\label{fig:t12_vs_t13}
\end{center}
\end{figure}

Analytically, we know that in of both models II and III, the mixing matrix has one constant column $\propto(2,-1,-1)$, corresponding to the TM1 mixing scheme, leading to well defined correlations between $\theta_{12}$ and $\theta_{13}$ (see Figure \ref{fig:t12_vs_t13})
\begin{eqnarray}
|U_{e1}|^2 = \cos^2\theta_{12} (1 - \sin^2\theta_{13}) = \frac{2}{3} \,.
\end{eqnarray}

\begin{figure}
\begin{center}
\includegraphics[scale=0.4]{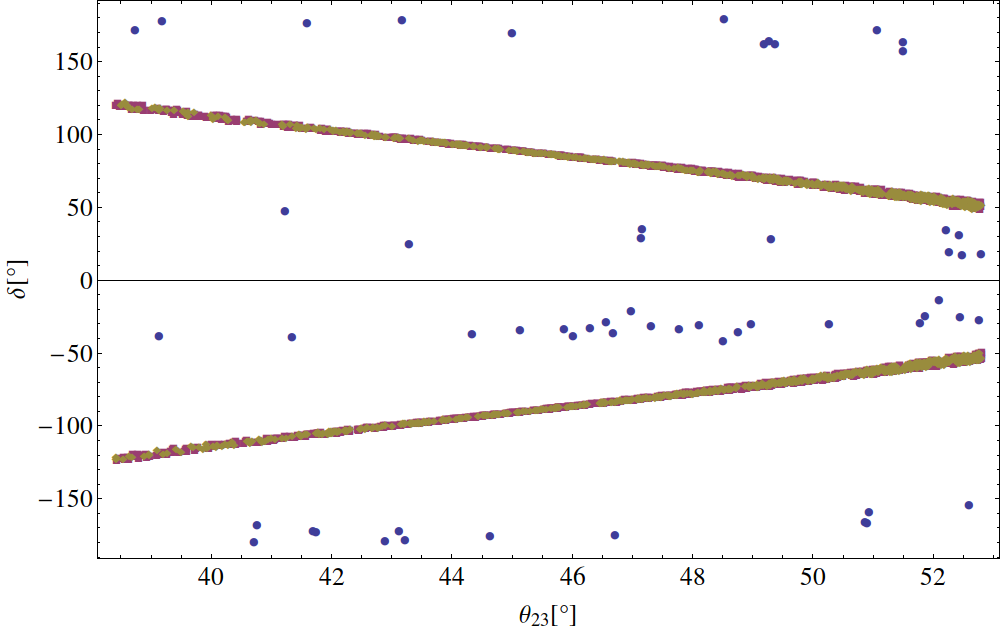}
\caption{Plot of the correlation between $\theta_{23}$ and $\delta$ in models I (blue circles), II (purple squares) and III (brown diamonds).}
\label{fig:t23_vs_delta}
\end{center}
\end{figure}

 Figure \ref{fig:t23_vs_delta} shows the predicted CP-violating phase $\delta$ correlated with $\theta_{23}$.
 Model I predicts CP violation but non maximal-violating value.
 Models II and III predict maximal CP violation at $\theta_{23}=45^\circ$.
 For $\theta_{23}$ deviating from the maximal mixing value, models II and III give the some correlation between $\delta$ and $\theta_{23}$ because of the TM1 mixing, 
\begin{eqnarray}
|U_{e2}|^2 = \sin^2\theta_{12} \cos^2\theta_{23} + \cos^2\theta_{12} \sin^2\theta_{23} \sin^2\theta_{13} + \frac{1}{2} \sin 2\theta_{12} \sin 2\theta_{23} \sin\theta_{13} \cos\delta= \frac{1}{6} \,.
\end{eqnarray}

\begin{figure}
\begin{center}
\includegraphics[scale=0.4]{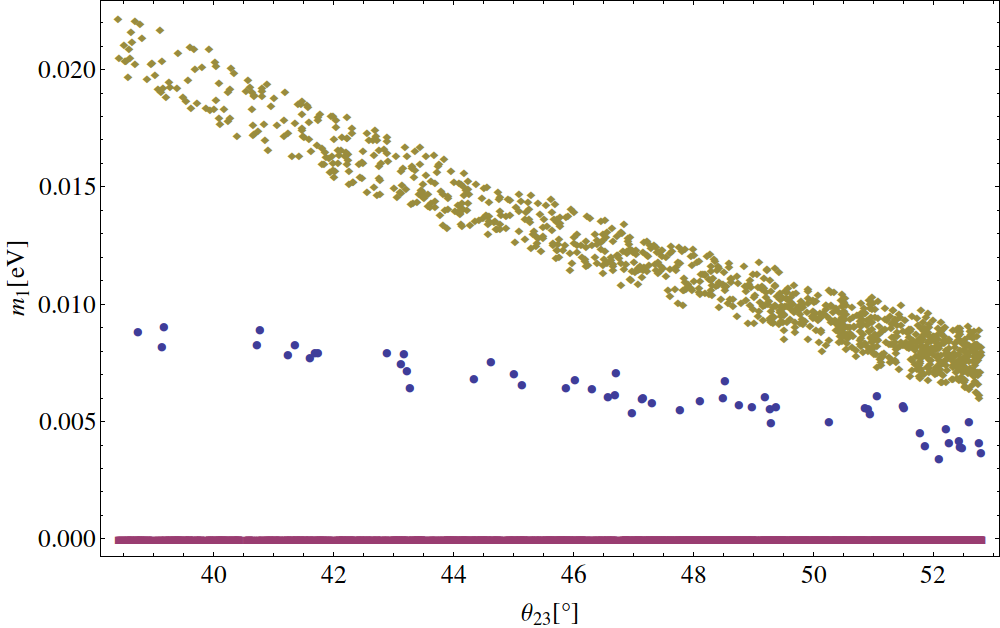}
\caption{Plot of the correlation between $\theta_{23}$ and $m_1$ in models I (blue circles), II (purple squares) and III (brown diamonds). Model II has $m_1=0$.}
\label{fig:t23_vs_m1}
\end{center}
\end{figure}

\begin{figure}
\begin{center}
\includegraphics[scale=0.4]{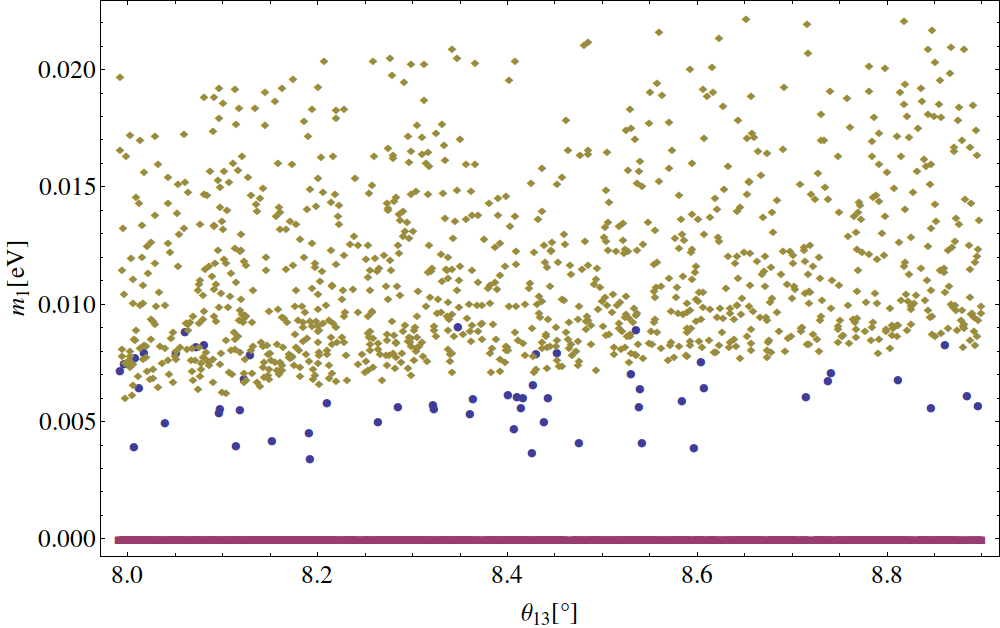}
\caption{Plot of the correlation between $\theta_{13}$ and $m_1$ in models I (blue circles), II (purple squares) and III (brown diamonds). Model II has $m_1=0$.}
\label{fig:t13_vs_m1}
\end{center}
\end{figure}

  In the remaining figures it is clear from the presence of some blue circles that Model I is also viable.
 This does not contradict the results of \cite{Bjorkeroth:2014vha}, where only the case with two RH neutrinos was excluded.
 In our parametrization, this corresponds to Model I, II or III with $\mu_1 = 0$. For $\mu_1 \neq 0$, Model II still predicts $m_1=0$. This is because the three vectors $V_1$, $V_2$ and $V_3$ are not linearly independent with each other in Model II and thus the rank of the mass matrix is 2. The three vectors $V_1$, $V_2$ and $V_3$ are linearly independent of each other, and thus predict three non-zero mass eigenvalues, which  can be verified that none of the viable points for Model I and III have $m_1 = 0$ (see e.g.
\ Figures \ref{fig:t23_vs_m1}, \ref{fig:t13_vs_m1}).

\begin{figure}
\begin{center}
\includegraphics[scale=0.4]{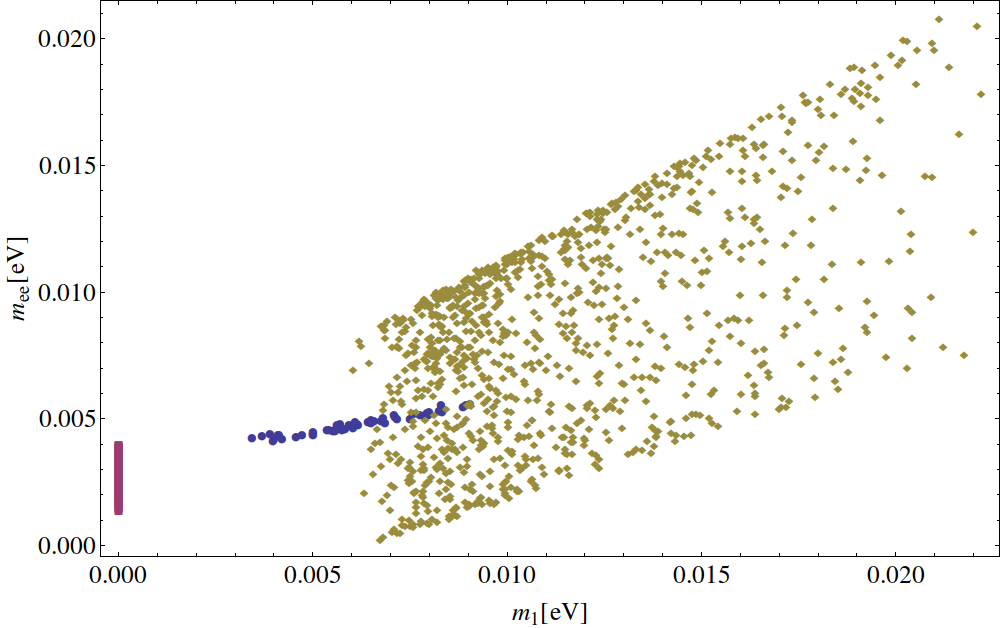}
\caption{Plot of the correlation between $m_1$ and $m_{ee}$ in models I (blue circles), II (purple squares) and III (brown diamonds). Model II has $m_1=0$. Note that the current best lower limit on $|m_{ee}|$ is at $0.18$ eV \cite{KamLAND-Zen:2016pfg}.}
\label{fig:m1_vs_mee}
\end{center}
\end{figure}

\begin{figure}
\begin{center}
\includegraphics[scale=0.4]{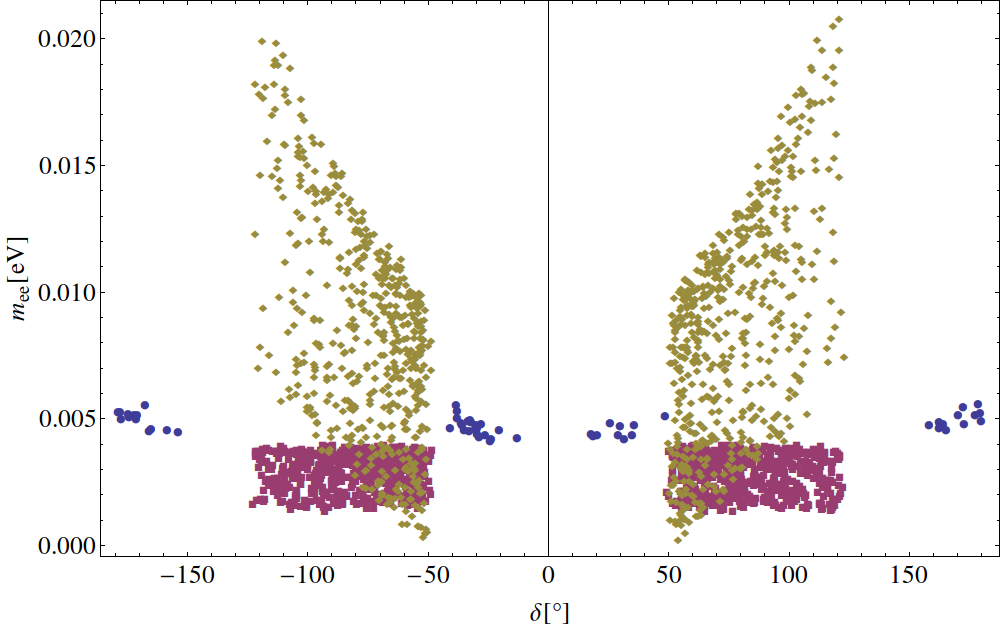}
\caption{Plot of the correlation between $\delta$ and $m_{ee}$ in models I (blue circles), II (purple squares) and III (brown diamonds). Model II has $m_1=0$. Note that the current best lower limit on $|m_{ee}|$ is at $0.18$ eV \cite{KamLAND-Zen:2016pfg}.}
\label{fig:delta_vs_mee}
\end{center}
\end{figure}

Figures \ref{fig:m1_vs_mee} and \ref{fig:delta_vs_mee} give correlations between the effective neutrinoless double beta decay parameter $m_{ee}$ and either $m_1$ or $\delta$.
 Figure \ref{fig:m1_vs_mee} shows a tight correlation and narrow allowed region for $m_{ee}$ in model I and for model II which can not be larger than about $0.
005$ eV (due to the smallness of $m_1$ in those models), whereas model III can have $m_{ee}$ going up to $0.
020$ eV as $m_1$ grows larger.

Figure \ref{fig:delta_vs_mee} gives the correlation between Dirac CP violating phase $\delta$ and $m_{ee}$.
 One can see $\delta$ has the same possible ranges for Model II and III (in accordance with both leading to TM1 mixing schemes).
 For Model I $\delta$ can take values of $\delta$ outside the ranges allowed by TM1 (when considering the experimental ranges of the angles).

%%%%

\section{Flavon cross couplings: their influences and absence\label{sec:cross_coupling}}

The total flavon potential should include not only potentials of each single flavon, but also cross couplings between them \cite{Pascoli:2016eld}. By introducing additional Abelian symmetries, parts of these couplings may be forbidden, but not all of them. There are always some cross couplings left.  These couplings may result in shifts of VEVs deviating from their original directions in Eq.~\eqref{eq:VEVs1}, affecting the results shown in the previous Section. In this section, we will analyse how cross couplings modify flavour mixing and provide a generic way to forbid them.

\subsection{Effects of cross couplings}

We take the following example to show how flavon VEVs are shifted and how the mixing is modified. The new Abelian symmetries we introduce are 
\bq
Z_3^l:&&(\mu_R,\Phi_l,\Phi_{l'})\to \omega (\mu_R,\Phi_l,\Phi_{l'})\,, \tau_R\to \omega^2 \tau_R \,,\nonumber\\
Z_2^{\nu'}:&&(N_3, \Phi_{\nu'})\to - (N_3, \Phi_{\nu'}) \,. 
\nq  
The only cross couplings that cannot be forbidden between two flavons $\Phi_A$ and $\Phi_B$ are 
\bq
V(\Phi_A,\Phi_B) = \sum_{\mathbf{r}} \epsilon^{AB}_\mathbf{r} \big( (\widetilde{\Phi}_A \Phi_A)_\mathbf{r} (\widetilde{\Phi}_B \Phi_B)_\mathbf{r} \big)_\mathbf{1}\,,
\nq
where $r=\mathbf{1}, \mathbf{2}, \mathbf{3}, \mathbf{3^\prime}$ and all coefficients $\epsilon^{AB}_\mathbf{r}$ are real required by the Hermitian of the Lagrangian. 
The whole flavon potential can be represented as
\bq
V = \sum_A V(\Phi_A) + \sum_{A\neq B} V(\Phi_A,\Phi_B) 
\label{eq:all_potential}
\nq
for $A, B$ sum for $\nu,\nu',l,l'$ and $A\neq B$. 
The shifted VEVs can be analytically obtained when assuming $|\epsilon^{AB}_\mathbf{r}|\ll |g_i|$. However, since there are too many cross coupling terms, the results are very complicated. On the other hand, as we only care about those that shift the directions of the flavon VEVs, it is unnecessary to account for all contributions for cross couplings. A simpler way to derive the flavon VEV shifts is to apply the residual symmetries of the VEVs at leading order. 
\begin{itemize}
\item
We first consider the residual symmetries that the VEVs in Eq.\ \eqref{eq:VEVs1} satisfy. $\langle \Phi_\nu \rangle$ is invariant under the actions of $S$ and $U$, namely satisfying a Klein symmetry $K_4=\,<\!S,U\!>$. $\langle \Phi_l \rangle$ is invariant under the actions of $T$ and $U$, satisfying a $S_3$ symmetry with $S_3=\,<\!T,U\!>$. Both $K_4$ and $S_3$ are subgroups of $S_4$. 
$\langle \Phi_{\nu'} \rangle$ does not satisfy any residual symmetries, but any products $\langle (\widetilde{\Phi}_{\nu'}\Phi_{\nu'})_\mathbf{r} \rangle$ satisfy the $K_4$ symmetry \footnote{We note that $\langle \Phi_{\nu'} \rangle $ preserves a ``square-root symmetry of $K_4$''.}. One can check that \footnote{Actually, $\langle (\widetilde{\Phi}_{\nu'}\Phi_{\nu'})_\mathbf{3^\prime} \rangle$ is invariant under the action of $-U$ not $U$ in the $\mathbf{3^\prime}$ representation, but the ``$-$'' sign does not matter since the $\mathbf{3^\prime}$ representation always appears in pairs in the flavon potential. }
\bq
\langle (\widetilde{\Phi}_{\nu'}\Phi_{\nu'})_\mathbf{2} \rangle \sim \begin{pmatrix}1 \\ 1 \end{pmatrix} \,,~ \langle (\widetilde{\Phi}_{\nu'}\Phi_{\nu'})_\mathbf{3} \rangle \sim 0 \,,~ 
\langle (\widetilde{\Phi}_{\nu'}\Phi_{\nu'})_\mathbf{3^\prime} \rangle \sim \begin{pmatrix}1 \\ 0 \\ 0 \end{pmatrix} \,. 
\nq
$\langle \Phi_{l'} \rangle$ does not satisfy any residual symmetries, either, but its products
\bq
\langle (\widetilde{\Phi}_{l'}\Phi_{l'})_\mathbf{2} \rangle \sim 0 \,,~ 
\langle (\widetilde{\Phi}_{l'}\Phi_{l'})_\mathbf{3} \rangle \sim 
\langle (\widetilde{\Phi}_{l'}\Phi_{l'})_\mathbf{3^\prime} \rangle \sim \begin{pmatrix}1 \\ 1 \\ 1 \end{pmatrix} \,
\nq  
satisfy the $S_3$ symmetries \footnote{We regard that $\langle \Phi_{l'} \rangle $ preserves a ``square-root symmetry of $S_3$''.}. 
Due to the above analysis, we conclude that all the combinations $\langle (\widetilde{\Phi}_A\Phi_A)_\mathbf{r} \rangle$ preserve $Z_2=\,<\!U\!>$ symmetry. 

\item
Another residual symmetry that we can use is the general CP (GCP) symmetry under the transformation 
\bq
\phi_1\leftrightarrow\phi_1^*\,, \quad \phi_2\leftrightarrow\phi_3^*\,.
\nq
This is an accidental symmetry that all VEVs in Eq.\ \eqref{eq:VEVs1} satisfy. One can also check that since each combination $(\widetilde{\Phi}_A \Phi_A)_\mathbf{r}$ corresponds to a flavour symmetry and the operator $\big( (\widetilde{\Phi}_A \Phi_A)_\mathbf{r} (\widetilde{\Phi}_B \Phi_B)_\mathbf{r} \big)_\mathbf{1}$ is invariant under this transformation, the potential $V$ is invariant under the CP transformation. 
\end{itemize}
Residual symmetries are powerful in deriving the structure of the modified flavon VEVs. Considering the correction to the VEVs $\langle \Phi_\nu \rangle$ and $\langle \Phi_l \rangle$, their VEVs preserve the $Z_2$ and GCP symmetries at leading order and every cross couplings also preserve them, thus 
\bq
\langle \Phi_\nu \rangle = \begin{pmatrix}1 \\ \epsilon_{\nu1} \\ \epsilon_{\nu1} \end{pmatrix} v_{\text{I}}\,,\quad
\langle \Phi_l \rangle = \begin{pmatrix}1+2\epsilon_{l1} \\ 1-\epsilon_{l1} \\ 1-\epsilon_{l1} \end{pmatrix} \frac{v_\text{III}}{\sqrt{3}}\,,
\label{eq:VEVs_corrections1}
\nq
where all parameters are real. To derive the correction to $\langle \Phi_{\nu'} \rangle$ and $\langle \Phi_{l'} \rangle$, one can naively express a CP-preserving VEV as
\bq
\langle \Phi_{\nu'} \rangle = \begin{pmatrix} \sqrt{2}\epsilon_{\nu2} \\ e^{+ i \pi/4}(1+i\epsilon_{\nu3}) \\  e^{- i \pi/4}(1-i\epsilon_{\nu3}) \end{pmatrix} \frac{v_\text{II}}{\sqrt{2}}\,,\quad
\langle \Phi_{l'} \rangle = \begin{pmatrix}1(1+2\epsilon_{l2}) \\ \omega(1-\epsilon_{l2}+i\sqrt{3}\epsilon_{l3}) \\ \omega^2(1-\epsilon_{l2}-i\sqrt{3}\epsilon_{l3}) \end{pmatrix} \frac{v_{\text{III}'}}{\sqrt{3}}\,,
\label{eq:VEVs_corrections2}
\nq
where all parameters are real. Corrections from all the other cross couplings will modify the sizes of $v_\text{I,II,III,III'}$, we can redefine them to absorb these effects, and after doing so, we arrive at the flavon VEVs in the form as in Eqs.\ \eqref{eq:VEVs_corrections1} and \eqref{eq:VEVs_corrections2}. 
In the second basis, these VEVs are transformed into 
\bq
&&\langle \Phi_\nu \rangle = \begin{pmatrix}1+2\epsilon_{\nu1} \\ 1-\epsilon_{\nu1} \\ 1-\epsilon_{\nu1} \end{pmatrix} \frac{v_\text{I}}{\sqrt{3}}\,,~
\langle \Phi_{\nu'} \rangle = \begin{pmatrix} \frac{1}{\sqrt{3}} + \frac{\epsilon_{\nu2}}{\sqrt{3}} -\frac{\epsilon_{\nu3}}{\sqrt{3}} \\ \frac{1}{3+\sqrt{3}} + \frac{\epsilon_{\nu2}}{\sqrt{3}} - \frac{-\epsilon_{\nu3}}{3-\sqrt{3}}  \\ \frac{-1}{3-\sqrt{3}} + \frac{\epsilon_{\nu2}}{\sqrt{3}} - \frac{\epsilon_{\nu3}}{3+\sqrt{3}}  \end{pmatrix} v_\text{II}\,,\nonumber\\
&&\langle \Phi_l \rangle = \begin{pmatrix}1 \\ \epsilon_{l1} \\ \epsilon_{l1} \end{pmatrix} v_{\text{III}}\,,\qquad\quad
\langle \Phi_{l'} \rangle = \begin{pmatrix} \epsilon_{l2}-\epsilon_{l3} \\ \epsilon_{l2} +\epsilon_{l3} \\ 1 \end{pmatrix} v_{\text{III}'}\,.
\label{eq:VEVs_corrections2}
\nq

For models with constant mixing patterns at leading order, we need the cross couplings to introduce next-to-leading order corrections to pull the mixing angles to the experimental allowed regimes \cite{Pascoli:2016eld}, such as models to realise TBM or TFH mixing. 

As an example of the possible effects of these perturbations on phenomenology, we reconsider the EAs in Eq.~\eqref{eq:unperturbed}
, which we then use to generate simplified models (with 2 RH neutrinos). These simplified models are not viable for the unperturbed EAs. The perturbed directions are:
\bq
\label{eq:perturbed}
V_1^\epsilon &=& \frac{1}{\sqrt{5}}\left[(1-\epsilon_{l1}) \begin{pmatrix} 1\\ 2 \\0 \end{pmatrix}  + \begin{pmatrix} - \epsilon_{l2} - \epsilon_{l3} - \epsilon_{\nu1}\\ +\epsilon_{l2} - \epsilon_{l3} + 4 \epsilon_{\nu1} \\ -3 \epsilon_{l2} - \epsilon_{l3} \end{pmatrix} \right] \,,
\\
V_2^\epsilon &=& \frac{1}{\sqrt{2}}\left[(1-\epsilon_{l1}-\epsilon_{\nu1}) \begin{pmatrix} 0\\1\\-1 \end{pmatrix} \right]\,. % + (0, - \epsilon_{l1} - \epsilon_{\nu1}, + \epsilon_{l1} + \epsilon_{\nu1})
\nq
It turns out that one of the EAs remains in the same direction after perturbations are taken into account, and we can just reabsorb the perturbation of $V_2^\epsilon$ into the overal proportional coefficient. This can only be done partially to $V_1^\epsilon$, from where we note also that the perturbation proportional to $\epsilon_{l3}$ is proportional to the EA (1,1,1) considered in the viable Model II, which indicates that the perturbed model with 2 RH neutrinos may be viable (for $\epsilon_{l2} = \epsilon_{\nu1}=0$ and sufficiently large $\epsilon_{l3}$ we would recover model II, although the $\epsilon_{l3}$ required may be too large to be consistent with $\epsilon_{l3}$ being a perturbation).

\begin{figure}
\begin{center}
\includegraphics[scale=0.4]{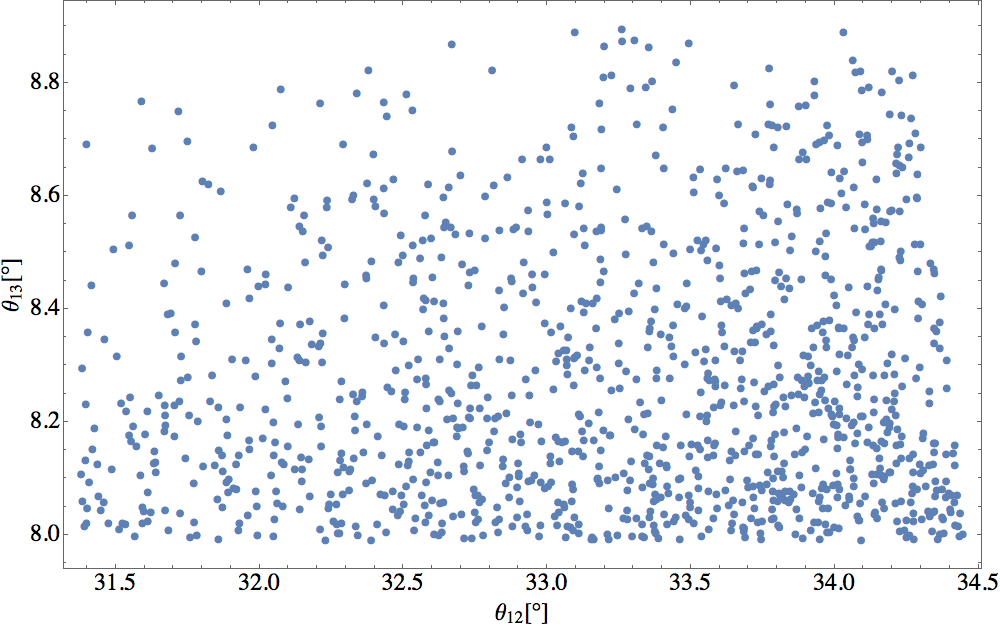}
\caption{Plot of the correlation between $\theta_{12}$ and $\theta_{13}$ in model with 2 RH neutrinos, including effects of cross-terms perturbing one of the EAs..}
\label{fig:t12_vs_t13_cross}
\end{center}
\end{figure}

\begin{figure}
\begin{center}
\includegraphics[scale=0.4]{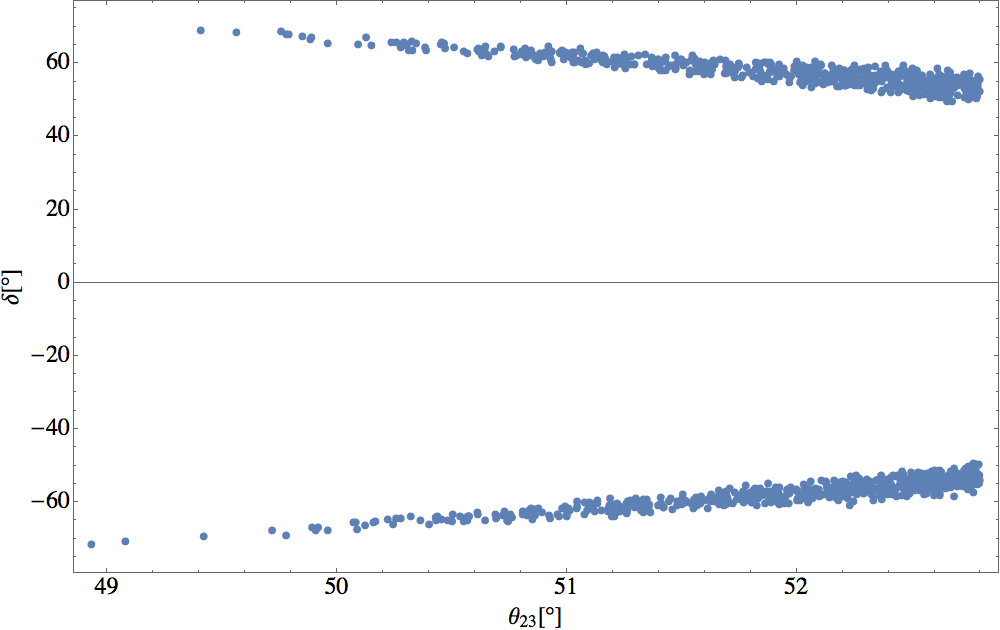}
\caption{Plot of the correlation between $\theta_{23}$ and $\delta$ in model with 2 RH neutrinos, including effects of cross-terms perturbing one of the EAs.}
\label{fig:t23_vs_delta_cross}
\end{center}
\end{figure}

The mass matrix in Eq.\eqref{eq:neutrino_mass} then becomes
\bq
M_\nu
= \sum \mu_i V_i^\epsilon V_i^{\epsilon T} \,,
\label{eq:neutrino_mass_perturbed}
\nq
and we consider the sum only with two EAs, as the model becomes viable even without the third RH neutrino.  Figs. \ref{fig:t12_vs_t13_cross} and \ref{fig:t23_vs_delta_cross} show points that are viable for all observables for this simplified model with 2 RH neutrinos, cf. Figs. \ref{fig:t12_vs_t13} and \ref{fig:t23_vs_delta}. Here, we take $\epsilon_{l2}$, $\epsilon_{l3}$ and $\epsilon_{\nu1}$ to vary randomly in the range $[0,0.1]$ for illustration.

We conclude that allowing for cross couplings increases the viability of models, at the potential cost of predictivity. In the example chosen, the 2 RH neutrino model becomes viable and is fairly predictive when comparing with the 3 RH neutrino models.

\subsection{Avoiding cross couplings}

For some other models, flavon cross couplings are unnecessary because they may make the model lose predictive power. In order to forbid these couplings, we need some other new physics at high scale.

Supersymmetry provides an option to avoid flavon cross couplings, although one should note the supersymmetry breaking soft terms reintroduce them, they are relatively supressed when compared with non-supersymmetric theories. We take the potential of a arbitrary $\varphi\sim \mathbf{3}$ for example. Introducing driving fields $\varphi^d_\mathbf{1}\sim\mathbf{1}$, $\varphi^d_\mathbf{2}\sim\mathbf{2}$, $\varphi^d_\mathbf{3} \sim\mathbf{3}$, $\varphi^d_\mathbf{3'}\sim\mathbf{3'}$, which take a $U(1)_R$ charge $2$, we construct the following superpotential
\bq
w= f_1' (\varphi\varphi)_{\mathbf{1}}\varphi^d_\mathbf{1} + f_2' (\varphi\varphi)_{\mathbf{2}}\varphi^d_\mathbf{2} + f_3' (\varphi\varphi)_{\mathbf{3}}\varphi^d_\mathbf{3} + f_4' (\varphi\varphi)_{\mathbf{3'}}\varphi^d_\mathbf{3'}.
\label{eq:superpotential}
\nq
The scalar potential is $V(\varphi)+V(\varphi^d)$, where
\bq
V(\varphi)=\sum_{\varphi^d_i} \left|\frac{\partial w}{\partial \varphi^d_i}\right|^2 + \mu_{\varphi}^2 ( \tilde{\varphi} \varphi )_\mathbf{1} \,,  \nonumber\\
V(\varphi^d)=\left|\frac{\partial w}{\partial \varphi}\right|^2 + \sum_{\varphi^d_i} \mu_{\varphi^d_i}^2 \left| \varphi^d_i \right|^2 \,. 
\label{eq:V1V2}
\nq 
Here, $\mu_{\varphi}^2$ and $\mu_{\varphi^d_i}^2$ are soft term mass parameters, which are much smaller than the supersymmetry scale. We do not need to care about $V(\varphi^d)$, because the driving fields only involve quadratic couplings in the potential and the minimum of $V(\varphi^d)$ is always zero at $\langle \varphi^d_i \rangle =0$. Then the minimisation of the total scalar potential is equivalent to that of $V(\varphi)$, which is exactly in Eq.~\eqref{eq:potential} with $f_{1,2,3,4}$ as functions of $f_{1,2,3,4}'$. Introducing less driving fields can provide additional constraints to the coefficients $f_{1,2,3,4}$. 

Applying this approach to the flavons $\Phi_l$, $\Phi_{l'}$, $\Phi_\nu$ and $\Phi_{\nu'}$, we construct the flavon potentials $V(\Phi_l)$, $V(\Phi_{l'})$, $V(\Phi_\nu)$ and $V(\Phi_{\nu'})$ from their superpotentials. Once we avoid to introduce cross couplings in the superpotential, no cross coupling in the flavon potential will be constructed in the flavon potential. 
In former works, such as in \cite{Altarelli:2005yx}, the minimisation of the flavon potential is always simplified to solving the following series of equations: 
\bq
\frac{\partial w}{\partial \varphi^d_i} = 0\,.
\label{eq:superpotential_min}
\nq 
This is valid in the limit of soft terms being much smaller than the supersymmetry scale, such that we can safely neglect the soft terms. 
Here in Eq.~\eqref{eq:superpotential}, however, as we have introduced redundant degrees of freedom for driving fields, the only solution for Eq.~\eqref{eq:superpotential_min} is $\langle \varphi^d_i\rangle = 0$. To find some non-trivial solutions, we have to return to minimise $V(\varphi)$ directly, as we did in section \ref{sec:VEV}. 

For UV-complete model constructions in the framework of supersymmetry, the holomorphy requirement does not allow couplings containing conjugates of any superfields in the superpotential. As a consequence, the required Yukawa structures via higher dimensional operators such as $(\tilde{\Phi}_\nu \Phi_l)_\mathbf{3} \overline{\ell_L} \tilde{H} N_2$ and $\big((\tilde{\Phi}_\nu \Phi_l)_\mathbf{3} \Phi_{l'}\big)_\mathbf{3} \overline{\ell_L} \tilde{H} N_3$ cannot be obtained. This situation is easily avoided by introducing extra flavon superfields which take the opposite $U(1)_F$ charges of $\Phi_l$ and $\Phi_\nu$ but all the other representation properties and the VEV directions the same as $\Phi_l$ and $\Phi_\nu$, respectively.

%%%%

%\newpage

\section{Conclusions}

In this paper we explored the construction of effective alignments and their use as building blocks for models. These effective alignments emerge in higher-order operators from the vacuum expectation values of flavon fields that arise naturally from the potentials of discrete flavour symmetries. Fermion mass terms are then built using such effective alignments and viable flavour models can be found.

This method can obtain certain alignments that could not be obtained otherwise. In certain cases, we obtain the same alignments obtained via other mechanisms in the literature, but in a much simpler way than aligning them directly. Some examples of this are the directions used in order to obtain the well-know tri-bimaximal mixing,  Toorop-Feruglio-Hagedorn mixing and in particular some of the directions of Constrained Sequential Dominance.

Considering potentials with 1 to 4 triplets of the flavour symmetry $S_4$, we classified the directions that are obtainable from the potential and then set out to construct models of the leptonic sector. We considered models where the charged lepton mass matrix is diagonal due to the flavour symmetry, and in the neutrino sector we employ 3 directions that are obtainable (either vacuum expectation values or effective alignments constructed from multiples of them).

For these models, we illustrate with renormalizable UV completions that models with effective alignments can be made predictive and viable, avoiding the proliferation of invariants that would have been allowed by the symmetries at the non-renormalizable level. We present three new viable models in the Constrained Sequential Dominance framework, all of which are compatible with neutrino oscillation data. Models II and III are new models that have not been considered before, they preserve TM1 mixing and predict almost maximal CP violation. Further to this, when cross terms are allowed in the potential, the simplified version of the model with only two right-handed neutrinos becomes viable.
Although we used the group $S_4$ explicitly, similar constructions and conclusions apply to other flavour symmetries.

\section*{Acknowledgements}

IdMV acknowledges funding from the Funda\c{c}\~{a}o para a Ci\^{e}ncia e a Tecnologia (FCT) through
the contract IF/00816/2015, partial support by Funda\c{c}\~ao para a Ci\^encia e a Tecnologia
(FCT, Portugal) through the project CFTP-FCT Unit 777 (UID/FIS/00777/2013) which
is partially funded through POCTI (FEDER), COMPETE, QREN and EU,
and partial support by the National Science Center, Poland, through the HARMONIA project under contract UMO-2015/18/M/ST2/00518.
The work of T.\,N. is supported by Spanish Grants No.
FPA2014-58183-P and No. SEV-2014-0398 (MINECO), and Grant No. PROMETEOII/2014/084
(Generalitat Valenciana). YLZ is funded by European Research Council under ERC Grant NuMass (FP7-IDEAS-ERC ERC-CG 617143). The authors would like to thank C.~Luhn, S.~King and S.~Pascoli for useful discussions. 

\appendix
\section{Group theory of $S_4$ \label{app:S4}} 

$S_4$ is the permutation group of 4 objects, see e.g.\ \cite{Escobar:2008vc}.
The Kronecker products between different irreducible representations can be easily obtained:
\begin{eqnarray}
&\mathbf{1^{\prime}}\otimes\mathbf{1^{\prime}}=\mathbf{1}, ~~\mathbf{1^{\prime}}\otimes\mathbf{2}=\mathbf{2}, ~~\mathbf{1^{\prime}}\otimes\mathbf{3}=\mathbf{3^{\prime}}, ~~
\mathbf{1^{\prime}}\otimes\mathbf{3^{\prime}}=\mathbf{3},~~\mathbf{2}\otimes\mathbf{2}=\mathbf{1}\oplus\mathbf{1}^{\prime}\oplus\mathbf{2},\nonumber\\
&
\mathbf{2}\otimes\mathbf{3}=\mathbf{2}\otimes\mathbf{3^{\prime}}=\mathbf{3}\oplus\mathbf{3}^{\prime},~~
\mathbf{3}\otimes\mathbf{3}=\mathbf{3^{\prime}}\otimes\mathbf{3^{\prime}}=\mathbf{1}\oplus \mathbf{2}\oplus\mathbf{3}\oplus\mathbf{3^{\prime}},~~
\mathbf{3}\otimes\mathbf{3^{\prime}}=\mathbf{1^{\prime}}\oplus \mathbf{2}\oplus\mathbf{3}\oplus\mathbf{3^{\prime}}
\end{eqnarray}
%

%%%%%%%%%%%%%%%
\begin{table}[h!]
\begin{center}
\begin{tabular}{cccc}
\hline\hline
   & $T$ & $S$ & U  \\\hline
$\mathbf{1}$ & 1 & 1 & 1 \\
$\mathbf{1^{\prime}}$ & 1 & 1 & -1 \\
$\mathbf{2}$ & 
$\left(
\begin{array}{cc}
 \omega  & 0 \\
 0 & \omega ^2 \\
\end{array}
\right)$ & 
$\left(
\begin{array}{cc}
 1 & 0 \\
 0 & 1 \\
\end{array}
\right)$ & 
$\left(
\begin{array}{cc}
 0 & 1 \\
 1 & 0 \\
\end{array}
\right)$ \\

$\mathbf{3}$ &  $\left(
\begin{array}{ccc}
 0 & 0 & 1 \\
 1 & 0 & 0 \\
 0 & 1 & 0 \\
\end{array}
\right)$ & 
$\left(
\begin{array}{ccc}
 1 & 0 & 0 \\
 0 & -1 & 0 \\
 0 & 0 & -1 \\
\end{array}
\right)$ &
$\left(
\begin{array}{ccc}
 1 & 0 & 0 \\
 0 & 0 & 1 \\
 0 & 1 & 0 \\
\end{array}
\right)$

\\

$\mathbf{3^{\prime}}$ &  
$\left(
\begin{array}{ccc}
 0 & 0 & 1 \\
 1 & 0 & 0 \\
 0 & 1 & 0 \\
\end{array}
\right)$ & 
$\left(
\begin{array}{ccc}
 1 & 0 & 0 \\
 0 & -1 & 0 \\
 0 & 0 & -1 \\
\end{array}
\right)$ &
$-\left(
\begin{array}{ccc}
 1 & 0 & 0 \\
 0 & 0 & 1 \\
 0 & 1 & 0 \\
\end{array}
\right)$  \\ \hline\hline

\end{tabular}
\caption{\label{tab:rep_matrix1} The representation matrices for the $S_4$ generators $T$, $S$ and $U$ in the first basis, where $\omega$ is the cube root of unit $\omega=e^{2\pi i/3}$.}
\end{center}
\end{table}
%%%%%%%%%%%%%%%%%%%%%%%%

In the main text, we have used two bases. This first one is helpful for deriving the full solutions of the flavon vacuum in the $S_4$ symmetry and the second one is applied when calculating flavour mixing. 
In the first basis, generators of $S_4$ in irreducible representations are given in Table \ref{tab:rep_matrix1}. In this basis, the products for two triplets $a=(a_1, a_2,a_3)^T$ and $b=(b_1, b_2, b_3)^T$ are divided into the following irreducible representations
\begin{eqnarray}
(ab)_\mathbf{1_i} &=& a_1b_1 + a_2b_2 + a_3b_3 \,,\nonumber\\
(ab)_\mathbf{2} &=& (
a_1b_1 + \omega a_2b_2 + \omega^2 a_3b_3 ,~
a_1b_1 + \omega^2 a_2b_2 + \omega a_3b_3 
)^T \,,\nonumber\\
(ab)_{\mathbf{3_i}} &=& \frac{1}{\sqrt{2}}(a_2b_3+a_3b_2, a_3b_1+a_1b_3, a_1b_2+a_2b_1)^T \,,\nonumber\\
(ab)_{\mathbf{3_j}} &=& \frac{i}{\sqrt{2}} (a_2b_3-a_3b_2, a_3b_1-a_1b_3, a_1b_2-a_2b_1)^T \,.
\label{eq:CG1}
\end{eqnarray} 
where 
\begin{eqnarray}
&&\mathbf{1_i}=\mathbf{1}\,, ~\; \mathbf{3_i}=\mathbf{3}\,, ~\; \mathbf{3_j}=\mathbf{3'}\,~\; \text{for} ~\; a\sim b \sim \mathbf{3}\,,~ \mathbf{3^\prime} \,, \nonumber\\
&&\mathbf{1_i}=\mathbf{1'}\,,~  \mathbf{3_i}=\mathbf{3'}\,,~ \mathbf{3_j}=\mathbf{3}\,~\;\; \text{for} ~\; a\sim  \mathbf{3}\,,~ b \sim \mathbf{3'}\,.
\end{eqnarray} 
And the products of two doublets $a=(a_1, a_2)^T$ and $b=(b_1, b_2)^T$ are divided into
\begin{eqnarray}
(ab)_\mathbf{1} &=& a_1b_2 + a_2b_1 \,,\quad
(ab)_\mathbf{1^\prime} = a_1b_2 - a_2b_1 \,,\quad
(ab)_{\mathbf{2}} = (a_2b_2, a_1b_1)^T \,,
\label{eq:CG_doublets}
\end{eqnarray} 
%%%%%%%%%%%%%%%
\begin{table}[h!]
\begin{center}
\begin{tabular}{cccc}
\hline\hline
   & $T$ & $S$ & U  \\\hline
$\mathbf{1}$ & 1 & 1 & 1 \\
$\mathbf{1^{\prime}}$ & 1 & 1 & -1 \\
$\mathbf{2}$ & 
$\left(
\begin{array}{cc}
 \omega  & 0 \\
 0 & \omega ^2 \\
\end{array}
\right)$ & 
$\left(
\begin{array}{cc}
 1 & 0 \\
 0 & 1 \\
\end{array}
\right)$ & 
$\left(
\begin{array}{cc}
 0 & 1 \\
 1 & 0 \\
\end{array}
\right)$ \\

$\mathbf{3}$ &  $\left(
\begin{array}{ccc}
 1 & 0 & 0 \\
 0 & \omega ^2 & 0 \\
 0 & 0 & \omega  \\
\end{array}
\right)$ &
$\frac{1}{3} \left(
\begin{array}{ccc}
 -1 & 2 & 2 \\
 2 & -1 & 2 \\
 2 & 2 & -1 \\
\end{array}
\right)$ &
$\left(
\begin{array}{ccc}
 1 & 0 & 0 \\
 0 & 0 & 1 \\
 0 & 1 & 0 \\
\end{array}
\right)$ \\

$\mathbf{3^{\prime}}$ &  $\left(
\begin{array}{ccc}
 1 & 0 & 0 \\
 0 & \omega ^2 & 0 \\
 0 & 0 & \omega  \\
\end{array}
\right)$ &
$\frac{1}{3} \left(
\begin{array}{ccc}
 -1 & 2 & 2 \\
 2 & -1 & 2 \\
 2 & 2 & -1 \\
\end{array}
\right)$ &
$-\left(
\begin{array}{ccc}
 1 & 0 & 0 \\
 0 & 0 & 1 \\
 0 & 1 & 0 \\
\end{array}
\right)$ \\ \hline\hline

\end{tabular}
\caption{\label{tab:rep_matrix2} The representation matrices for the $S_4$ generators $T$, $S$ and $U$ in the second basis, where $\omega$ is the cube root of unit $\omega=e^{2\pi i/3}$.}
\end{center}
\end{table}
%%%%%%%%%%%%%%%%%%%%%%%%
% 
The generators of $S_4$ in the second basis in different irreducible representations are listed in Table \ref{tab:rep_matrix2}.  
This basis is widely used in the literature since the charged lepton mass matrix invariant under $T$ is diagonal in this basis. The products of two 3 dimensional irreducible representations $a$ and $b$ can be expressed as
\begin{eqnarray}
(ab)_\mathbf{1_i} &=& a_1b_1 + a_2b_3 + a_3b_2 \,,\nonumber\\
(ab)_\mathbf{2} &=& (a_3b_3 + a_1b_2 + a_2b_1,~ a_2b_2 + a_1b_3 + a_3b_1)^T \,,\nonumber\\
(ab)_{\mathbf{3_i}} &=& \frac{1}{\sqrt{6}} (2a_1b_1-a_2b_3-a_3b_2, 2a_3b_3-a_1b_2-a_2b_1, 2a_2b_2-a_3b_1-a_1b_3)^T \,,\nonumber\\
(ab)_{\mathbf{3_j}} &=& \frac{1}{\sqrt{2}} (a_2b_3-a_3b_2, a_1b_2-a_2b_1, a_3b_1-a_1b_3)^T \,.
\label{eq:CG2}
\end{eqnarray}
The products of two doublets stay the same as in Eq.\ \eqref{eq:CG_doublets} in the first basis. 

The Kronecker products of multiplets of $S_4$ require the following properties: if the trilinear combination of three multiplets $a\sim \mathbf{r}$, $b\sim \mathbf{r'}$ and $c\sim \mathbf{r''}$ is an invariance of $S_4$, e.g., $\big((ab)_\mathbf{r''}c\big)_\mathbf{1}$, then equation
\begin{eqnarray}
\big((ab)_{\mathbf{r}_c}c\big)_\mathbf{1}\,=\big((bc)_{\mathbf{r}_a}a\big)_\mathbf{1}\;=\big((ca)_{\mathbf{r}_b}b\big)_\mathbf{1} \,
=\big(a(bc)_{\mathbf{r}_a}\big)_\mathbf{1}\;=\big(b(ca)_{\mathbf{r}_b}\big)_\mathbf{1}\, =\big(c(ab)_{\mathbf{r}_c}\big)_\mathbf{1}\,\,, \nonumber\\
\big((ab)_{\mathbf{r}'_c}c\big)_\mathbf{1'}=\big((bc)_{\mathbf{r}'_a}a\big)_\mathbf{1'}=\big((ca)_{\mathbf{r}'_b}b\big)_\mathbf{1'} 
=\big(a(bc)_{\mathbf{r}'_a}\big)_\mathbf{1'}=\big(b(ca)_{\mathbf{r}'_b}\big)_\mathbf{1'} =\big(c(ab)_{\mathbf{r}'_c}\big)_\mathbf{1'}\,.
\end{eqnarray}
holds, where $\mathbf{r}'=\mathbf{1'},\mathbf{1},\mathbf{2},\mathbf{3'},\mathbf{3}$ for $\mathbf{r}=\mathbf{1},\mathbf{1'},\mathbf{2},\mathbf{3},\mathbf{3'}$, respectively.
The above equation can be proved by expanding the Kronecker products explicitly. For example, using the Clebsch-Gordan coefficients in Eq.\ \eqref{eq:CG1}, we derive 
\begin{eqnarray}
\big((ab)_\mathbf{3}c\big)_\mathbf{1}=\sum_{i,j,k=1,2,3}\frac{1}{\sqrt{2}}|\epsilon_{ijk}|a_i b_j c_k = \sum_{i,j,k=1,2,3}\frac{1}{\sqrt{2}}|\epsilon_{jki}|b_j c_k a_i = \big((bc)_\mathbf{3}a\big)_\mathbf{1}\,, 
\label{eq:invariance}
\end{eqnarray}
for $a \sim b \sim c \sim \mathbf{3}$, and
\begin{eqnarray}
\big((ab)_\mathbf{3'}c\big)_\mathbf{1}=\sum_{i,j,k=1,2,3}\frac{i}{\sqrt{2}}\epsilon_{ijk}a_i b_j c_k = \sum_{i,j,k=1,2,3}\frac{i}{\sqrt{2}}\epsilon_{jki}b_j c_k a_i = \big((bc)_\mathbf{3}a\big)_\mathbf{1}\,, 
\label{eq:invariance_p}
\end{eqnarray}
for $a \sim b \sim \mathbf{3}$ and $c \sim \mathbf{3'}$.

\section{Minimisation of the $S_4\times U(1)_F$ potential \label{app:min}}

To find the possible VEVs of $\varphi$, we minimize $V(\varphi)$. One of the necessary conditions for the minimum of $V(\varphi)$ is
\begin{eqnarray}
\varphi_i^* \frac{\partial V(\varphi)}{\partial \varphi_i^*}\Big|_{\varphi=\langle \varphi \rangle} &=& \frac{v_i^2}{2} \Big[ \mu_\varphi^2 + g_1 v_i^2 + \big( g_1+\frac{1}{2}g_2 \big) (v_j^2+v_k^2) + \frac{1}{2} g_3 \big( v_j^2 e^{2 i \alpha_{ji}} +  v_k^2 e^{2 i \alpha_{ki}} \big) \Big] =0  \,,
\label{eq:key}
\end{eqnarray}
where $\alpha_{ji}=\alpha_j-\alpha_i$, and $ijk=123,231,312$. 
A VEV must be (meta-)stable, which requires the second derivative of $V(\varphi)$ to be positive at this value. In other words, the matrix $M^2$ with entries $M_{ij}^2$ defined through
\bq
M_{ij}^2=\frac{\partial^2 v(\varphi)}{\partial \varphi_i^* \partial \varphi_j}\Big|_{\varphi=\langle \varphi \rangle} \,
\nq
must be positive-definite. 
We distinguish the solutions into three classes based on if $v_i$ vanishes. 
\begin{itemize}
\item {Case I: One of $v_i$ is non-zero, while the others are vanishing.} Without loss of generality, we assume $v_1$ is non-zero and $v_2=v_3=0$. 
Eq.\ \eqref{eq:key} is simplified to 
$\mu_\varphi^2 + g_1 v_1^2 =0$, 
and the solution is given by  $v_1^2={-\mu_\varphi^2}/{g_1}$
and $V(\varphi)$ at this value is given by
$V(\varphi)_\text{I}=-{\mu_\varphi^4}/{(4g_1)}$. 
The phase $\alpha_1$ cannot be determined. $M^2$ at $\langle \varphi \rangle_\text{I}$ is diagonal, with non-vanishing values 
\bq
m_1^2=-2\mu_\varphi^2\,, \quad m_2^2=m_3^2=\frac{g_2+g_3}{2g_1} (-\mu_\varphi^2) \,.
\nq
To make $m_2^2$ and $m_3^2$ positive, we must require $g_2+g_3>0$. 

\item {Case II: Two $v_i$ are non-zero, while the other one is vanishing.} 
Without lose of generality, we assume $v_2$ and $v_3$ are non-zero and $v_1=0$.
Eq.\ \eqref{eq:key} is simplified to
\begin{eqnarray}
\mu_\varphi^2 + g_1 v_2^2 + (g_1+\frac{1}{2}g_2) v_3^2 + \frac{1}{2}g_3 v_3^2 e^{2i\alpha_{32}} =0\,. \nonumber\\
\mu_\varphi^2 + g_1 v_3^2 + (g_1+\frac{1}{2}g_2) v_2^2 + \frac{1}{2}g_3 v_2^2 e^{2i\alpha_{23}} =0\,.
\end{eqnarray}
Since all the coefficients are real, the imaginary part of the above equation should be vanishing, which leads to $\alpha_{32} =n\pi/2$ with $n=0,1,2,3$. The real part of the equation leads to
\bq
v_2^2=v_3^2=\frac{-2\mu_\varphi^2}{4g_1+g_2+\eta g_3}\,
\nq
with $\eta=+1$ for $n=0,2$ and $\eta=-1$ for $n=1,3$, respectively. 
$V(\varphi)$ at this value is given by
\bq
V(\varphi)_\text{II}=-\frac{\mu_\varphi^4}{4g_1+g_2+\eta g_3}\,.
\nq
The eigenvalues of $M^2$ in this case are given by 
\bq
&&m_1^2=-2\mu_\varphi^2\,, \quad -m_2^2=2m_3^2=\frac{2(g_2+g_3)}{4g_1+g_2+g_3} (-\mu_\varphi^2) \,;\nonumber\\
&&m_1^2=-2\mu_\varphi^2\,, \quad m_2^2=\frac{2(g_3-g_2)}{4g_1+g_2-g_3} (-\mu_\varphi^2)\,,\quad m_3^2=\frac{g_2+g_3}{4g_1+g_2-g_3} (-\mu_\varphi^2) \,
\nq
for $\eta=\pm1$, respectively. Note that since $m_2^2$ and $m_3^2$ cannot take both positive values for $\eta=1$, the first class corresponds to a saddle point of $V(\varphi)$ and cannot be a vacuum. Thus, we only keep the other case, $\eta=-1$, or equivalently, $n=1,3$. The requirement of positive $m_2^2$ and $m_3^2$ is $g_3>0$ and $-g_3<g_2<g_3$.

\item {Case III: All $v_i$ do not vanish. }
Eq.\ \eqref{eq:key} is simplified to
\bq
&&\mu_\varphi^2 + g_1 v_1^2 + (g_1+\frac{1}{2}g_2) (v_2^2+v_3^2)
+\frac{1}{2} g_3 (v_2^2 e^{2 i \alpha_{21}} + v_3^2 e^{2 i \alpha_{31}}) =0 \,,\nonumber\\
&&\mu_\varphi^2 + g_1 v_2^2 + (g_1+\frac{1}{2}g_2) (v_3^2+v_1^2) 
+\frac{1}{2} g_3 (v_2^2 e^{2 i \alpha_{32}} + v_3^2 e^{2 i \alpha_{12}}) =0 \,,\nonumber\\
&&\mu_\varphi^2 + g_1 v_3^2 + (g_1+\frac{1}{2}g_2) (v_1^2+v_2^2) 
+\frac{1}{2} g_3 (v_2^2 e^{2 i \alpha_{13}} + v_3^2 e^{2 i \alpha_{23}}) =0 \,.
\label{eq:caseIII}
\nq
There are two classes of solutions: 
\bq
2\alpha_{12}= 2\alpha_{13}=0,\,\quad &&
v_1^2=v_2^2=v_3^2= \frac{-\mu_\varphi^2}{3g_1+g_2+g_3} \,;\nonumber\\
2 \alpha_{12} = - 2 \alpha_{13}=\frac{2}{3}\pi,\,\frac{4}{3}\pi\,,\quad &&
v_1^2=v_2^2=v_3^2= \frac{-\mu_\varphi^2}{3g_1+g_2-g_3/2} \,.
\nq
The corresponding value of $V(\varphi)$ is 
\bq
V(\varphi)_\text{III}= - \frac{3\mu_\varphi^4}{4(3g_1+g_2+g_3)} \,,\quad V(\varphi)_{\text{III}'}=- \frac{3\mu_\varphi^4}{4(3g_1+g_2-g_3/2)} \,.
\nq
The eigenvalues of $M^2$ in this case are given by 
\bq
&&m_1^2=-2\mu_\varphi^2\,, \quad m_2^2=\frac{-(g_2+g_3)}{3g_1+g_2+g_3} (-\mu_\varphi^2)
\,, \quad m_2^2=\frac{-3g_3}{3g_1+g_2+g_3} (-\mu_\varphi^2) \,;\nonumber\\
&&m_1^2=-2\mu_\varphi^2\,, \quad m_{2,3}^2=\frac{2g_3-g_2\pm\sqrt{g_2^2+2g_2g_3+10 g_3^2}}{6g_1+2g_2-g_3} (-\mu_\varphi^2) \,
\nq
for $\eta=\pm1$, respectively. In the case $\eta=+1$, the requirements of positive eigenvalues are $g_3<0$ and $g_2+g_3<0$. In the other case, the requirements are $g_3>0$ and $g_2+g_3<0$. 
\end{itemize}

In summary, each of these VEVs can be obtained when the following conditions hold.
\begin{itemize}
\item
If $g_2+g_3>0$ and $g_2-g_3>0$, the only possible VEV for $\varphi$ is $\langle \varphi \rangle_\text{I}$ because $V(\varphi)$ at $\langle \varphi \rangle_\text{I}$ is the only local minimum of the potential and thus also the global minimum. 

\item
If $g_2+g_3>0$ and $g_2-g_3<0$, both $\langle \varphi \rangle_\text{I}$ and $\langle \varphi \rangle_\text{II}$ could be the vacua of $\varphi$. The flavon potential has two classes of local minimums, at $\langle \varphi \rangle_\text{II}$ and $\langle \varphi \rangle_\text{III}$, respectively, and that at $\langle \varphi \rangle_\text{II}$ is the global one. For random values of the parameters, $\varphi$ has a larger chance to gain a VEV at $\langle \varphi \rangle_\text{II}$.

\item
If $g_2+g_3<0$ and $g_3<0$, $V(\varphi)$ at $\langle \varphi \rangle_\text{III}$ is the only local and thus the global minimum of the flavon potential. $\langle \varphi \rangle_\text{III}$ is the only choice of vacuum of $\varphi$. 

\item
If $g_2+g_3<0$ and $g_3>0$, $V(\varphi)$ at $\langle \varphi \rangle_{\text{III}'}$ is the only local and thus the global minimum of the flavon potential. $\langle \varphi \rangle_{\text{III}'}$ is the only choice of vacuum of $\varphi$. 

\end{itemize}

\end{document}